\documentclass[journal]{IEEEtran}

\usepackage{cite}
\usepackage{graphicx}
\usepackage{amsmath}
\usepackage{amssymb}
\usepackage{algorithm}
\usepackage{algorithmic}
\usepackage{array}
\usepackage{url}
\usepackage{booktabs}
\usepackage{multirow}
\usepackage{caption}
\usepackage{subcaption}
\usepackage{xcolor}

\UseRawInputEncoding
\begin{document}
\title{Active Perception for Radio Map Reconstruction in Uncharted 3D Air-Ground Environments}

\author{Wenlihan Lu, \IEEEmembership{Graduate Student Member, IEEE}, Miaowen Wen, \IEEEmembership{Senior Member, IEEE}, and Shijian Gao, \IEEEmembership{Member, IEEE}
    \thanks{Part of work has been presented at 2025 IEEE International Conference on Wireless Communications and Signal Processing\cite{Lu2025Activeradiomap}. 
    
    Wenlihan Lu and Shijian Gao are with the Internet of Things, The Hong Kong University of Science and Technology (Guangzhou), China (email: wlu162@connect.hkust-gz.edu.cn; shijiangao@hkust-gz.edu.cn). 
    Miaowen Wen is with the School of Electronic and Information Engineering, South China University of Technology, Guangzhou, China (email: eemwwen@scut.edu.cn).}
}

\markboth{Journal of \LaTeX\ Class Files,~Vol.~14, No.~8, August~2015}%
{Shell \MakeLowercase{\textit{et al.}}: Bare Demo of IEEEtran.cls for IEEE Journals}

\maketitle
\newcommand{\dist}[2]{\left\lVert #1-#2 \right\rVert_2}
\newcommand{\Eall}{\mathbb{E}_{\mathcal{D},\mathbf{M}}}
\begin{abstract}
    Radio maps provide the essential foundation for low altitude networking systems. Unlike terrestrial radio maps that are typically generated via drive test measurements, mapping the air-ground environment requires the deployment of unmanned aerial vehicles (UAVs). This shift introduces two formidable challenges in uncharted 3D scenarios. First, sparse radio measurements and incomplete geometric observations hinder accurate reconstruction. Second, the large 3D action space and strict power constraints from high spectrum scanner energy consumption make informative exploration difficult. To address these issues, this paper proposes \textbf{{3D}} \textbf{{u}}ncertainty \textbf{{a}}ware \textbf{{r}}adio \textbf{{a}}ctive \textbf{{m}}apping (\textbf{3D-URAM}), a closed-loop active perception framework that decouples the mapping process into two offline trained stages. In Stage I, a Bayesian U-Net is developed to recover radio maps from sparse measurements and partial geometry while providing predictive uncertainty. In Stage II, a dynamic probabilistic roadmap and a Transformer-based waypoint selection policy trained via proximal policy optimization maximize long horizon uncertainty reduction under travel budgets. Experimental results demonstrate that 3D-URAM reduces reconstruction error by over 50\% compared to representative baselines. Real-world field tests within a 300m$\times$200m space also validate the potential of active radio map reconstruction.
\end{abstract}

\IEEEpeerreviewmaketitle

\section{Introduction}

Emerging 6G mobile computing applications require wireless networks to evolve from reactive communication to proactive, environment aware intelligence. Radio maps provide a digital representation of the physical radio environment by characterizing the spatial variation of wireless propagation over a region of interest~\cite{Zeng2024ckm, Wu2018Radiomap, Majeed2016Radiomap}. It can encode key channel metrics, including reference signal received power (RSRP), which is widely used to reflect large scale signal coverage and link quality~\cite{Chikha2024REMMIMO,Zhang2019Cellular,Perz2015REMinterferencemanagement}. Such spatial intelligence is particularly crucial for emerging low altitude wireless networks~\cite{Gao2026ISAC}, where reliable UAV connectivity requires informed resource management and trajectory design~\cite{Li2024Large}. Moreover, radio maps can provide spatial context for DRL-based resource coordination and beamforming~\cite{Yuanjian2023RRM}, as well as outage-aware path planning~\cite{Yuanjian2022CCUAV}.
Despite its importance, constructing accurate radio maps in 3D low altitude environments remains a formidable challenge~\cite{Romero2022Radiomapsurvey}. Unlike two dimensional terrestrial networks, aerial signal propagation is intrinsically three dimensional and highly heterogeneous, frequently subjected to severe shadowing and multipath fading within complex urban canyons~\cite{Hu20233dradiomap}. Deterministic ray tracing (RT) models~\cite{Jakob2023Sionna} provide physically rigorous predictions, but they rely on exhaustive and up to date geometric priors, which are often unavailable in practical deployments~\cite{Bakirtzis2025RMDeeplearning}.

To mitigate RT's complete geometric knowledge and heavy computation limitations, extensive efforts have explored alternative reconstruction paradigms. Statistical interpolation techniques, including kriging~\cite{Sato2017Kriging},  matrix completion~\cite{Sun2022Matrixcompletion} and Gaussian processes (GP)~\cite{Noel1993GP}, exploit spatial correlations but oversimplify the complex radio propagation patterns.
Recently, deep learning has become the dominant approach for high fidelity radio map reconstruction~\cite{Teganya2022Autoencoder}. Representative deep learning methods, including RadioUNet~\cite{Levie2021radiounet}, GAN based models~\cite{Zhang2023RMEGAN}, graph neural network (GNN) based methods~\cite{Chen2023GNNRM}, Transformer-based architectures~\cite{Viet2025Transformer, Zhao2025Spectrum}, and diffusion based frameworks~\cite{Luo2025Denoiseradiomap,Wang2025radiodiff,gao2026farmfoundationalaerialradio}, improve both reconstruction fidelity and spatial realism. Beyond deterministic prediction, \cite{Krijestorac2021Spatial} employ a U-Net to output the mean and variance of a Gaussian, providing an early probabilistic treatment of radio map estimation.

Despite their good performance in certain scenarios, these reconstruction advances fundamentally rely on the availability of pre-collected and sufficiently dense datasets. A crucial challenge lies in how such informative measurements are acquired. In realistic deployments, establishing a massive network of static sensors to guarantee dense coverage is difficult, making mobile agents like UAVs a preferred alternative to sample radio signals. Recent UAV-aided 3D spectrum mapping research likewise highlights efficient sampling under limited resources as a key design requirement~\cite{jie2026UAV3DRM}. However, UAV-borne measurement is constrained by strict power consumption budgets, induced by both the UAV's limited battery capacity and the energy required by the carried onboard spectrum scanner, inevitably rendering sparse observations~\cite{Shrestha2023Spectrumsurveying}. This reality necessitates active radio mapping: the agent autonomously orchestrates its exploration to acquire the most informative measurements under limited power budgets, deeply coupling map estimation with sequential path planning.

\begin{table*}[t]
    \centering
    \caption{Comparison of Radio Map Reconstruction and Active Radio Mapping.}
    \label{tab:research_comparison}
    \renewcommand{\arraystretch}{1.15}
    \setlength{\tabcolsep}{2.5pt}
    \scriptsize
    \newcommand{\yesmark}{\ensuremath{\checkmark}}
    \newcommand{\nomark}{\ensuremath{\times}}
    \newcommand{\partialmark}{\ensuremath{\triangle}}
    \begin{tabular}{@{}>{\centering\arraybackslash}p{0.09\textwidth}
                        >{\centering\arraybackslash}p{0.20\textwidth}
                        >{\centering\arraybackslash}p{0.10\textwidth}
                        >{\centering\arraybackslash}p{0.11\textwidth}
                        >{\centering\arraybackslash}p{0.07\textwidth}
                        >{\centering\arraybackslash}p{0.21\textwidth}
                        >{\centering\arraybackslash}p{0.16\textwidth}@{}}
        \toprule
        \textbf{Refs.} &
        \textbf{Method} &
        \textbf{3D radio map} &
        \textbf{Geometric prior} &
        \textbf{TX prior} &
        \textbf{Uncertainty quantification} &
        \textbf{Active planning} \\
        \midrule
        \cite{Sato2017Kriging,Sun2022Matrixcompletion,Noel1993GP} &
        Statistical interpolation &
        \nomark & \nomark & \nomark & Posterior var. (GP only) & \nomark \\
        \cite{Teganya2022Autoencoder,Levie2021radiounet,Zhang2023RMEGAN,Chen2023GNNRM,Viet2025Transformer,Zhao2025Spectrum,Luo2025Denoiseradiomap,Wang2025radiodiff,gao2026farmfoundationalaerialradio} &
        Deterministic deep learning &
        \partialmark & Full & \partialmark & \nomark & \nomark \\
        \cite{Krijestorac2021Spatial} &
        Probabilistic deep learning &
        \nomark & Full & \yesmark & Aleatoric only & \nomark \\
        \cite{Polyzos2024ActiveGP,Chen2026IPPGP} &
        Active gaussian process &
        \partialmark & \nomark & \nomark & Posterior variance & Short-horizon \\
        \cite{Shrestha2023Spectrumsurveying} &
        Active post-hoc uncertainty &
        \nomark & Full & \nomark & Post-hoc & Short-horizon \\
        \cite{Matson2024Oreman} &
        Active post-hoc uncertainty &
        \nomark & Partial & \nomark & Post-hoc & Short-horizon \\
        \textbf{This work} &
        \textbf{3D-URAM (proposed)} &
        \textbf{\yesmark} & \textbf{Partial} & \textbf{\nomark} &
        \textbf{Aleatoric + epistemic} &
        \textbf{Long-horizon RL policy} \\
        \bottomrule
    \end{tabular}
    \vspace{1mm}

    {\scriptsize\textit{Legend:} \yesmark: included; \partialmark: partially included; \nomark: not included.\par}
\end{table*}

Existing mobile agent based active radio mapping methods attempt to address the coupling between reconstruction and exploration by guiding sampling with predictive uncertainty or heuristic next-best-viewpoint criteria. Early studies use GP to jointly model signal distribution and predictive variance~\cite{Polyzos2024ActiveGP,Chen2026IPPGP}, but they suffer from scalability limitations and insufficient representational power in complex environments. Other approaches rely on neural predictors with post-hoc uncertainty estimators~\cite{Shrestha2023Spectrumsurveying,Matson2024Oreman}, which can produce poorly calibrated confidence in unseen regions. More importantly, existing methods remain limited in uncharted 3D airspace for two reasons: they typically rely on short-horizon greedy planners~\cite{Bircher2016NBV} that cannot effectively balance long-term information gain against energy budgets, and they often assume known free space or static geographic databases that fail to capture transient obstructions~\cite{OpenStreetMap,Jaensch2026aerialimage}. As a result, they are not well suited to active sensing in realistic 3D environments with only partial geometric priors. Table~\ref{tab:research_comparison} summarizes these research lines.

To solve the challenges of partial geometric prior, obstacle-aware navigation and long-horizon active sensing, we propose 3D uncertainty aware radio active mapping (3D-URAM), a closed-loop framework that tightly integrates online geometric perception, uncertainty aware radio reconstruction, and budget constrained long horizon exploration in uncharted 3D environments. In Stage~I, a Bayesian U-Net trained with dual masked inpainting recovers a dense 3D radio map from sparse measurements and partial geometry, while quantifying predictive uncertainty to identify where additional measurements are valuable. In Stage~II, a reinforcement learning policy operates on an online constructed dynamic probabilistic roadmap (PRM)~\cite{Vashisth2024dprm} to select obstacle free waypoints that maximize long horizon uncertainty reduction under travel budget. The two modules are trained offline and composed into an iterative perception action loop for autonomous 3D radio map acquisition.
The main contributions of this paper are summarized as follows:
\begin{itemize}
    \item A 3D Bayesian U-Net is built to address the radio map reconstruction under sparse measurements and incomplete geometric priors. This approach is powerful in its ability to recover signal fields while simultaneously quantifying both aleatoric and epistemic uncertainty.
    \item A 3D dynamic probabilistic mapping is proposed to handle large volumetric action spaces in unknown environments. This component enables the incremental construction of a collision free waypoint graph online directly from partial geometric perception.
    \item A Transformer-based waypoint selection policy is crafted to maximize the long horizon uncertainty reduction. This policy represents a significant advancement by optimizing information gain under strict travel budgets.
    \item Extensive evaluations on a large scale benchmark demonstrate that 3D-URAM achieves higher reconstruction accuracy than representative baselines. A supplementary open-loop field case study in a 300m$\times$200m space confirms its practical potential for real world~applications. 
\end{itemize}

The remainder of this paper is organized as follows. Section II formulates the problem. Sections III and IV detail the proposed Stage I reconstruction and Stage II active exploration, respectively. Section IV also presents closed-loop deployment and complexity analysis. Section V presents the experimental results, and Section VI concludes the paper. Notations are listed in Table~\ref{tab:notation}.

\begin{table}[t]
    \centering
    \caption{Notation Table}
    \label{tab:notation}
    \renewcommand{\arraystretch}{1.2}
    \small
    \begin{tabular}{l >{\raggedright\arraybackslash}p{0.69\linewidth}}
        \toprule
        \textbf{Variables} & \textbf{Definition} \\
        \midrule
        $\Omega$ & 3D region of interest. \\
        $\mathbf{p}_k$ & UAV position at decision step $k$. \\
        $\mathcal{C}(\mathbf{p}_k)$ & Camera field-of-view coverage region at $\mathbf{p}_k$. \\
        $\mathcal{M}_k$ & Global occupancy map fused up to step $k$. \\
        $\mathcal{W}_k$ & Candidate waypoint set at step $k$. \\
        $\mathbf{w}_{k,j}$ & $j$-th candidate waypoint in $\mathcal{W}_k$. \\
        $\Gamma(\mathbf{p}_k,\mathbf{p}_{k+1})$ & Vertical-then-horizontal flight route for transition $k$. \\
        $\Psi_k$ & Radio measurements collected along $\Gamma(\mathbf{p}_k,\mathbf{p}_{k+1})$. \\
        $\mathcal{Y}_k$ & Cumulative radio measurements available at step $k$. \\
        $P,\ \hat{P}$ & Ground-truth and reconstructed radio map (received power field). \\
        $f_\theta$ & Reconstruction network with parameters $\theta$. \\
        $\pi$ & Sensing (waypoint-selection) policy. \\
        $c(\mathbf{p}_k,\mathbf{p}_{k+1})$ & Route length of one waypoint transition. \\
        $C(\mathcal{T}_\pi)$ & Cumulative route length of trajectory $\mathcal{T}_\pi$. \\
        $B$ & Maximum cumulative route-length budget. \\
        \bottomrule
    \end{tabular}
\end{table}

\section{Problem Statement}\label{sec:system_model}
This section establishes the system model for 3D active radio map construction and formulates the optimization of reconstruction and exploration under UAV travel budget.
\subsection{System Model}\label{subsec:models}
\textbf{Environmental Map Construction.}
Consider an autonomous UAV operating within an unknown 3D region $\Omega$. We discretize $\Omega$ into a 3D voxel grid and represent the environment using an occupancy map for navigation and radio map reconstruction. Let $\mathbf{p}_k = [x_k, y_k, z_k]^\top \in \Omega$ denote the UAV's position at discrete decision step $k$. The UAV's onboard camera provides real-time geometric perception: at each step $k$, it observes a local region $\mathcal{C}(\mathbf{p}_k) \subset \Omega$ and estimates a height map of surrounding structures within the field of view (FoV).

\begin{figure}
    \centering
    \includegraphics[width=1\linewidth]{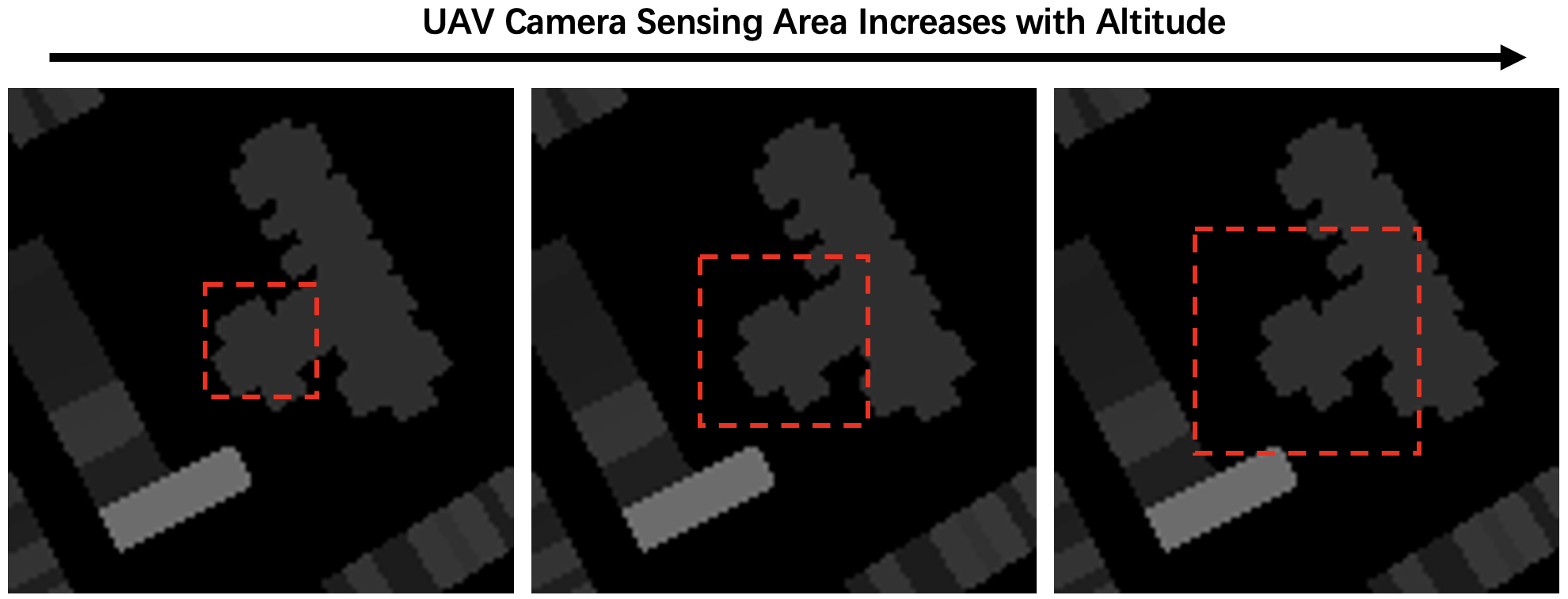}
    \caption{UAV camera FoV at different altitudes.}
    \label{fig:placeholder}
\end{figure}

The scope of visual perception is intrinsically coupled with the UAV's flight altitude. Let $\psi_{\mathrm{cam}}$ denote the camera aperture angle and $z_k$ the altitude at step $k$. The coverage region $\mathcal{C}(\mathbf{p}_k)$ expands with altitude and can be approximated as a pyramidal projection whose lateral extent scales with $z_k \cdot \tan(\psi_{\mathrm{cam}}/2)$; thus, a higher flight altitude enables the agent to capture a broader spatial area within a single observation as shown in Fig.~\ref{fig:placeholder}.
To support radio map reconstruction and safe navigation, the agent maintains a voxelized occupancy map $\mathcal{M}_k$ inferred from camera based geometric perception. At each step $k$, the estimated height map in $\mathcal{C}(\mathbf{p}_k)$ is voxelized into a local occupancy grid; voxels corresponding to buildings and obstacles are marked as occupied and fused into the global map $\mathcal{M}_k$, which persists across steps and provides geometric information for radio map reconstruction.
\[
    \mathcal{M}_k = \bigcup_{i=0}^{k} \mathcal{C}(\mathbf{p}_i).
\]

At each step $k$, the agent generates a fixed size set of candidate waypoints $\mathcal{W}_k = \{\mathbf{w}_{k,1}, \mathbf{w}_{k,2}, \dots, \mathbf{w}_{k,m}\}$ by sampling from the obstacle free portion of the current camera view. The UAV follows a vertical-then-horizontal motion rule between consecutive waypoints. For any target $\mathbf{w}=[x_w,y_w,z_w]^\top$, define the intermediate point $\mathbf{q}(\mathbf{p}_k,\mathbf{w})=[x_k,y_k,z_w]^\top$ and the resulting two-leg route
\[
    \Gamma(\mathbf{p}_k,\mathbf{w})
    \triangleq
    \operatorname{seg}\!\left(\mathbf{p}_k,\mathbf{q}(\mathbf{p}_k,\mathbf{w})\right)
    \cup
    \operatorname{seg}\!\left(\mathbf{q}(\mathbf{p}_k,\mathbf{w}),\mathbf{w}\right).
\]
Thus, the UAV first ascends or descends vertically to the target altitude and then flies horizontally to the target waypoint. A candidate $\mathbf{w}_{k,j}$ is valid only if both legs of $\Gamma(\mathbf{p}_k,\mathbf{w}_{k,j})$ are free of obstacles and do not intersect occupied voxels in $\mathcal{M}_k$.

\textbf{Radio Map Sampling.}
The radio map is assumed time invariant over the acquisition period. The radio map is characterized by a reference received signal power field $P: \Omega \rightarrow \mathbb{R}$. In addition to the camera, the UAV carries a lightweight spectrum scanner that continuously acquires signal measurements along its flight path.

At step $k$, the sampling policy selects an optimal target waypoint $\mathbf{w}_k^\star \in \mathcal{W}_k$, and the UAV position is updated as $\mathbf{p}_{k+1} = \mathbf{w}_k^\star$. During the transition along the vertical-then-horizontal route $\Gamma(\mathbf{p}_k,\mathbf{p}_{k+1})$, the scanner continuously collects observations, formally defined as:
\[
    \Psi_k \triangleq \bigl\{ \bigl(\mathbf{x}, P(\mathbf{x}) \bigr) \mid \mathbf{x} \in \Gamma(\mathbf{p}_k,\mathbf{p}_{k+1}) \bigr\},
\]
The cumulative radio sampling available for radio map reconstruction up to step $T$ is given by the union of all transition routes: $\mathcal{Y}_T = \bigcup_{k=0}^{T-1} \Psi_k$. For convenience, we denote the cumulative measurements available at step $k$ as $\mathcal{Y}_k = \bigcup_{i=0}^{k-1} \Psi_i$.

\subsection{Problem Formulation}\label{subsec:problem_formulation}
Given the cumulative measurements $\mathcal{Y}_T$ and the incrementally built environmental map $\mathcal{M}_T$ at the final step $T$, a neural estimator $f_\theta$ reconstructs the global radio map:
\[
    \hat{P} = f_\theta\bigl(\mathcal{Y}_T, \mathcal{M}_T\bigr).
\]
Let $e=(P^e,\mathcal{O}^e)\sim\mathcal{D}$ denote an environment instance comprising its ground-truth radio map and building geometry. Our goal is to minimize the expected reconstruction error, measured by a metric $\mathcal{E}(\cdot,\cdot)$ such as the mean squared error (MSE), over both the region of interest and the environment distribution. Because the environment and the radio map are initially unknown, this requires jointly learning the reconstruction parameters $\theta$ and the sensing policy $\pi$ that maps the current agent state to the next waypoint $\mathbf{p}_{k+1}$. The joint optimization problem is formulated as:
\begin{equation}
    \begin{aligned}
        \min_{\theta,\,\pi}\quad & \mathbb{E}_{e\sim\mathcal{D}}\,\mathbb{E}_{\mathcal{T}_\pi \sim p_\pi(\cdot\mid e)}\Bigl[\mathcal{E}\Bigl(P^e, f_\theta(\mathcal{Y}^e_T, \mathcal{M}^e_T)\Bigr)\Bigr] \\[2pt]
        \text{s.t.}\quad         & C(\mathcal{T}_\pi) \le B,
    \end{aligned}
    \label{eq:problem}
\end{equation}
where $p_\pi(\cdot\mid e)$ denotes the trajectory distribution induced by $\pi$ in environment $e$, $\mathcal{T}_\pi = \{\mathbf{p}_0, \mathbf{p}_1, \dots, \mathbf{p}_T\}$, and $(\mathcal{Y}^e_T,\mathcal{M}^e_T)$ are the cumulative measurements and environmental map obtained along that trajectory. We define the cumulative travel cost as $C(\mathcal{T}_{\pi}) \triangleq \sum_{k=0}^{T-1} c(\mathbf{p}_k,\mathbf{p}_{k+1})$. Under the adopted vertical-then-horizontal motion protocol, the transition cost is
\[
    c(\mathbf{p}_k,\mathbf{p}_{k+1})
    = |z_{k+1}-z_k|
    + \left\|[x_{k+1}-x_k,\,y_{k+1}-y_k]^\top\right\|_2.
\]
Thus, $B$ specifies the maximum cumulative route length available for exploration. The remaining budget is initialized as $b_0=B$ and evolves according to $b_{k+1}=b_k-c(\mathbf{p}_k,\mathbf{p}_{k+1})$. At each decision step, candidates satisfying $c(\mathbf{p}_k,\mathbf{w}_{k,i})\le b_k$ form the feasible action set. For the approximately constant-speed operating regime considered here, the route length provides an efficient proxy for propulsion expenditure. When platform-specific power profiles are available, direction-dependent propulsion, hovering, sensing, payload, and communication costs can be incorporated into $c(\cdot,\cdot)$ without altering the policy-training procedure~\cite{Yuanjian2025Mec}.

Equation~\eqref{eq:problem} couples two interdependent components: the reconstruction network $f_\theta$, which defines a belief over the radio map from sparse measurements, and the sensing policy $\pi$, which determines future measurements under the travel budget. Directly optimizing this problem end-to-end is highly intractable because (i) the environment and radio map are only partially observable during exploration, requiring the policy to act on a $\theta$-dependent belief state;  (ii) the loop requires expensive on-policy rollouts and non-differentiable operations such as waypoint sampling, graph reconstruction, and collision checking, leading to high sample complexity.

\subsection{Solution Overview: Two Stage Decoupled Framework}\label{subsec:two_stage}
\begin{figure*}[t]
    \centering
    \includegraphics[width=0.9\linewidth]{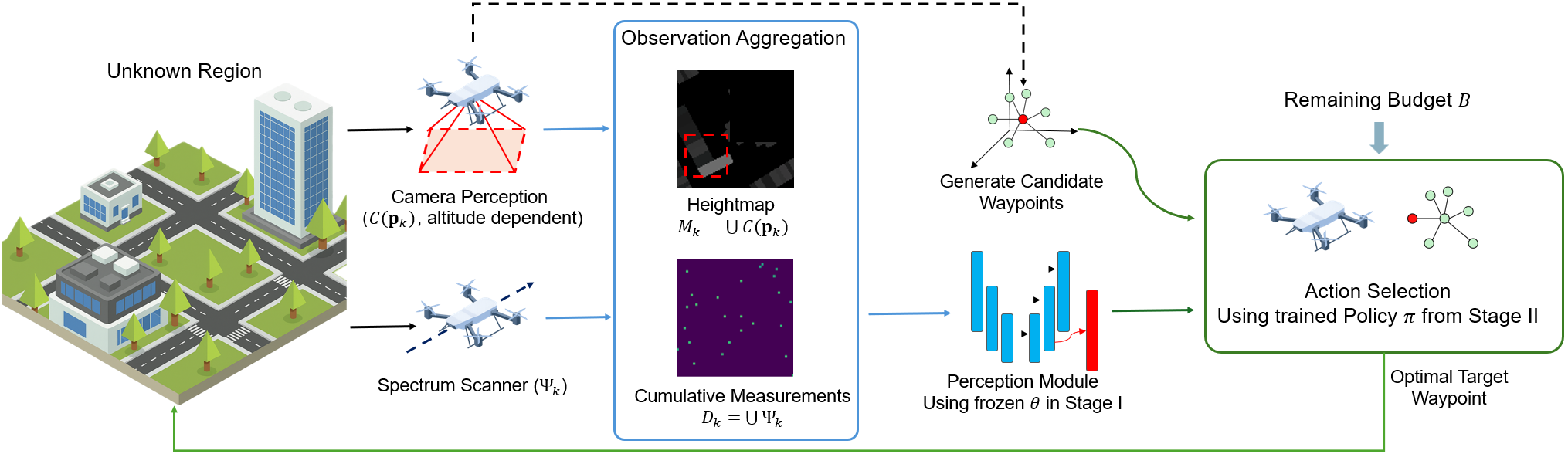}
    \caption{3D-URAM overview: Stage~I trains a Bayesian reconstructor to output a radio map belief from sparse measurements and partial geometry; Stage~II uses a dynamic PRM and a Transformer-based policy to select informative collision free waypoints under a travel budget.}
    \label{fig:framework}
\end{figure*}

To address these challenges, a two stage decoupling is adopted, comprising two offline training stages followed by an online closed-loop deployment, as summarized in Fig.~\ref{fig:framework}.

\begin{enumerate}
    \item \textbf{Stage~I:} The reconstruction parameters $\theta$ are optimized under dense supervision with dual masked inpainting, yielding a Bayesian predictor $f_{\theta^\star}$ that outputs the posterior mean radio map and predictive uncertainty, as detailed in Sec.~\ref{sec:perception}. The uncertainty map provides an intrinsic reward signal, serving as a computable surrogate for the ground-truth reconstruction error in Eq.~\eqref{eq:problem}, which is unavailable online.
    \item \textbf{Stage~II:} With $\theta^\star$ frozen, the sensing policy $\pi$ is trained via reinforcement learning to select collision free waypoints that maximize long horizon reduction of the predicted uncertainty under the travel budget, as detailed in Sec.~\ref{sec:drl_method}.
\end{enumerate}
This decomposition ensures that exploration is guided by a predictive belief state, while the reconstruction network remains agnostic to the specific exploration strategy.

\section{Stage I: Uncertainty-Aware Radio Map Reconstruction}\label{sec:perception}
In active radio mapping, the autonomous agent must explicitly identify poorly reconstructed regions to guide subsequent measurements. Consequently, Stage~I constructs a Bayesian reconstruction architecture that recovers radio map while simultaneously quantifying prediction uncertainty.

\subsection{Observation Modeling via Dual Masked Mechanism}
While existing approaches to radio map reconstruction typically assume that the complete environmental geometry is known, they rarely account for the complex, multimodal sensing patterns encountered in uncharted environments. In practical active mapping, the building layout is initially unknown; a deployed agent must incrementally perceive the geometry and infer $P$ from partial observations collected. More importantly, these practical observations arrive through two sensing modalities with fundamentally different spatial structures: the spectrum scanner acquires received radio power samples along the UAV trajectory, yielding sparse, scattered point like radio signals on the 2D grid, whereas the onboard camera reveals building geometry only within its FoV, producing block like contiguous regions of valid geometric priors. By neglecting the partial observability of environmental geometry and the structural dichotomy between point wise signal sampling and block wise visual sensing, conventional generic masking schemes fail to faithfully emulate the complex observation missingness characteristic of actual deployment.

To explicitly model this heterogeneous missingness, we generate two factorized masks during offline training:
\begin{itemize}
    \item \textbf{Sparse radio mask.} The radio mask $\mathbf{M}_r \in \{0,1\}^{H \times W}$ is sampled as independent and identically distributed random variables on each grid cell as $[\mathbf{M}_r]_{u,v} \sim \mathrm{Bernoulli}(p_r)$, which follows the conventional random sampling based masking approaches in radio map reconstruction. The retention probability $p_r$ controls the expected measurement density and therefore the difficulty of the inpainting problem.
    \item \textbf{Block wise geometric mask.} To emulate FoV limited perception, the height map $\mathcal{H}_k$ (rasterized from $\mathcal{M}_k$) is partitioned into non overlapping $s{\times}s$ patches $\{\mathcal{P}_j\}$. Each patch is retained with probability $p_g$ by sampling $g_j\sim\mathrm{Bernoulli}(p_g)$ and setting $[\mathbf{M}_g]_{u,v}=g_j$ for all $(u,v)\in\mathcal{P}_j$. In practice, we randomize $p_g$ within a range to cover varying visible ratios caused by viewpoint and occlusion.
\end{itemize}

During early online exploration, the acquired radio channel observations can be extremely sparse. Directly training the reconstruction network with a near zero radio mask $\mathbf{M}_r$ from scratch often exacerbates the ill posed nature of the inpainting problem, leading to severe optimization instability and unstable uncertainty estimates. To mitigate this, a curriculum learning schedule is adopted where the retention probability $p_r$ is linearly annealed over the training epochs. Specifically, the training initiates with a relatively dense masking regime. By initially providing a larger number of samples, the network can more effectively capture the underlying spatial correlations between the observed sample points and the overall radio environment. As training progresses, $p_r$ is progressively decremented to approach the target extreme sparsity regime encountered during actual deployment. This staged, monotonically increasing training difficulty effectively stabilizes convergence and substantially enhances both reconstruction robustness and uncertainty awareness under severe observation scarcity.

Given the cumulative observations $(\mathcal{Y}_k,\mathcal{M}_k)$ at decision step $k$, we rasterize them into fixed resolution grids. Let $\mathbf{P}^{\mathrm{obs}}_k \triangleq \mathcal{R}_P(\mathcal{Y}_k) \in \mathbb{R}^{H\times W}$ denote the sparse radio observation grid (unobserved cells filled with zero), and let $\mathcal{H}_k \triangleq \mathcal{R}_H(\mathcal{M}_k) \in \mathbb{R}^{H\times W}$ denote the rasterized height map. The network input $\mathbf{X}_k$ concatenates the rasterized radio observations and the partially observed height map:
\begin{equation}
    \mathbf{X}_k = \text{Concat}(\mathbf{P}^{\mathrm{obs}}_k, \mathcal{H}_k \odot \mathbf{M}_g),
    \label{eq:input_concat}
\end{equation}
where $\odot$ denotes the Hadamard product. During offline training, $\mathbf{P}^{\mathrm{obs}}_k$ is simulated via $\mathbf{P}^{\mathrm{obs}}_k = P \odot \mathbf{M}_r$, while during online deployment it is directly rasterized from the cumulative measurements $\mathcal{Y}_k$.

\subsection{Bayesian U-Net Architecture}
Building upon the success of prior studies~\cite{Lu2025sip2net}, the U-Net architecture is adopted as the backbone for radio map reconstruction due to its proven empirical performance and rapid inference speed. The efficacy of U-Net in this domain primarily stems from its encoder--decoder structure augmented with skip connections. The encoder efficiently extracts multiscale spatial features from the raw inputs, capturing both local propagation phenomena and macroscopic pathloss trends. Concurrently, the skip connections preserve high resolution spatial details by directly routing information to the decoder, enabling the network to accurately recover fine grained structural variations such as shadow boundaries and diffraction effects introduced by the complex urban topology.

\begin{figure}[t]
    \centering
    \includegraphics[width=1.0\linewidth]{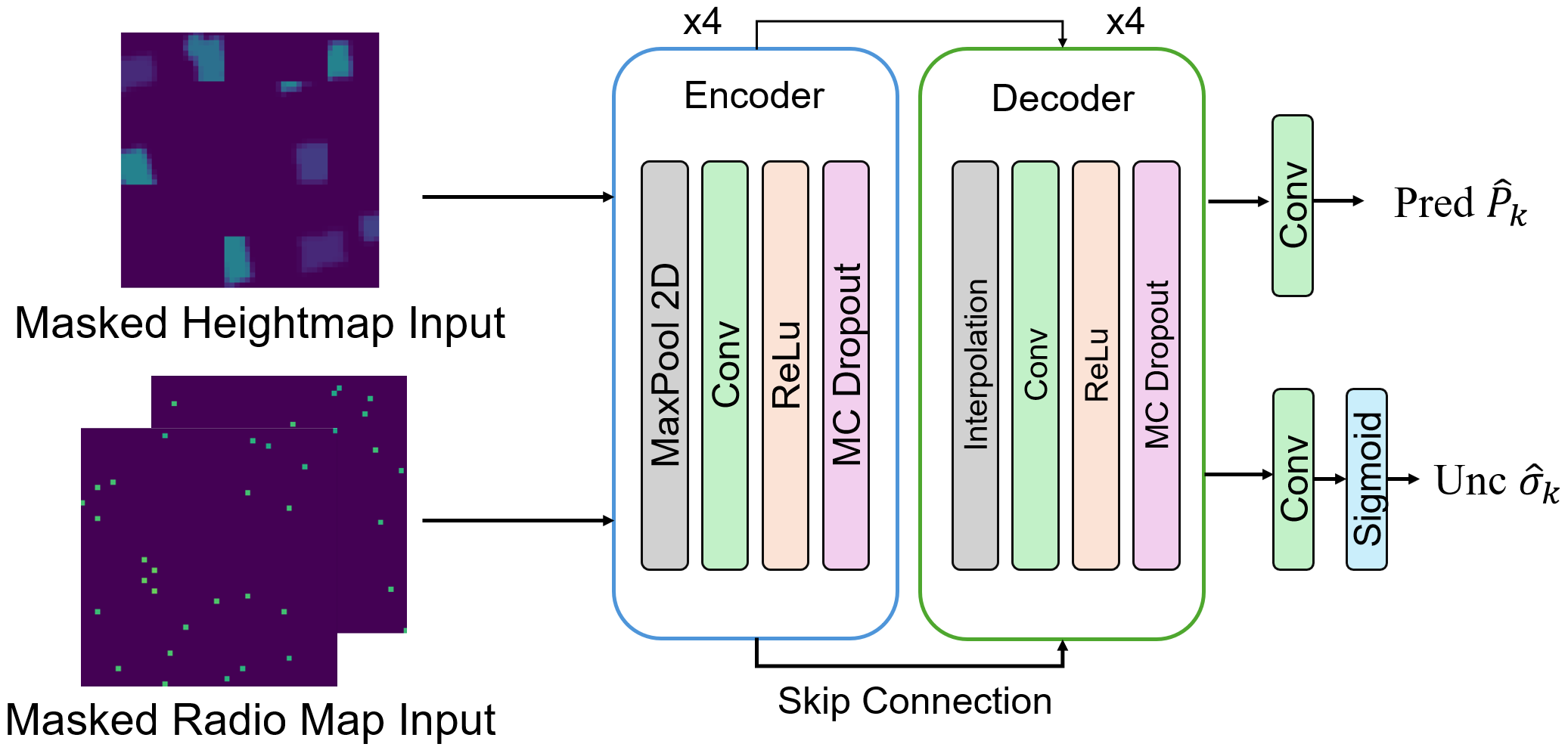}
    \caption{Bayesian U-Net for map reconstruction and uncertainty estimation.}
    \label{fig:net_arch}
\end{figure}

As detailed in Fig.~\ref{fig:net_arch}, the proposed Bayesian U-Net processes the concatenated masked height map and masked radio map inputs through a symmetric four stage encoder--decoder architecture. Each of the four encoder stages consists of a 2D Max Pooling layer for spatial downsampling, a Convolutional layer, a rectified linear unit (ReLU) activation function, and a Monte Carlo (MC) Dropout layer. Symmetrically, each of the four decoder stages employs an Interpolation layer for spatial upsampling, followed by a Conv layer, a ReLU activation, and an MC Dropout layer. Skip connections are utilized to directly route intermediate multiscale feature maps from the encoder to the corresponding decoder stages.

At the end of the decoder, the network uses two output heads: one predicts the mean signal strength $\hat{P}_k$, and the other predicts heteroscedastic variance $\hat{\sigma}_k^2$. Together with the MC Dropout layers, this structural design transforms the reconstructor into a Bayesian predictor $f_\theta$. During inference, it aggregates stochastic samples to output a posterior mean radio map $\bar{P}_k$ and a predictive uncertainty map $U_k$. The theoretical formulation of this joint uncertainty quantification is detailed in the subsequent section.

\subsection{Bayesian Uncertainty Quantification}
Uncertainty in radio map reconstruction arises from two complementary sources: aleatoric uncertainty due to inherent channel randomness, fading and measurement noise, and epistemic uncertainty due to model ignorance under sparse and biased sampling. We quantify both within a unified Bayesian framework and generate a belief state that can be consumed by downstream active exploration.

\subsubsection{Aleatoric Uncertainty via Heteroscedastic Negative Log Likelihood (NLL) Loss}
Observation noise in practical radio propagation is inherently \emph{spatially non-stationary}. In open line-of-sight (LoS) regions, macroscopic attenuation is primarily governed by smooth, distance dependent decay, resulting in high predictability and consequently bounded noise levels. Conversely, in the vicinity of dense building clusters, particularly around non-line-of-sight (NLoS) transitions, complex physical phenomena, such as diffraction, multifaceted scattering, and shadow fading, induce rapid, high frequency signal fluctuations. These stochastic variations cannot be deterministically resolved from partial geometric priors alone. By explicitly estimating a heteroscedastic variance map $\hat{\sigma}_i^2$, the proposed architecture is physically grounded to differentiate between intrinsically predictable areas and highly volatile environments, thereby capturing the spatial heterogeneity of aleatoric uncertainty with high fidelity. This motivates a spatially varying noise model, implemented via the variance head.

Leveraging the dual head output, the decoder departs from conventional deterministic map reconstruction by predicting a mean--variance pair $(\hat{P}_i, \hat{\sigma}_i^2)$ for each spatial grid cell $i$. During the training stage, the complete radio map $P$ serves as the dense ground-truth target, while the dual masks simulate the severe observation missingness encountered during inference. To explicitly account for the spatially non-stationary observation noise, the network is optimized over the environment distribution and sampled dual masks by minimizing a heteroscedastic negative log-likelihood loss across the set of valid grid cells $\mathcal{Q}$. Here, $\Eall$ abbreviates the joint expectation $\mathbb{E}_{(P,\mathcal{H})\sim\mathcal{D}}\mathbb{E}_{\mathbf{M}_r,\mathbf{M}_g}$:
\begin{equation}
    \begin{aligned}
        \mathcal{L}_{\mathrm{nll}}(\boldsymbol{\theta})
        &= \mathbb{E}_{(P,\mathcal{H})\sim\mathcal{D}}\,
        \mathbb{E}_{\mathbf{M}_r,\mathbf{M}_g}
        \left[-\frac{1}{|\mathcal{Q}|}\sum_{i\in\mathcal{Q}}
        \ln p\!\left(P_i\mid\hat{P}_i,\hat{\sigma}_i^2\right)\right] \\
        &= \Eall\Biggl[
        \frac{1}{|\mathcal{Q}|}\sum_{i\in\mathcal{Q}}
        \left(\frac{(P_i-\hat{P}_i)^2}{2\hat{\sigma}_i^2}
        +\frac{1}{2}\log\!\left(2\pi\hat{\sigma}_i^2\right)\right)\Biggr].
    \end{aligned}
    \label{eq:nll}
\end{equation}
In this formulation, the squared residual term is adaptively attenuated by the predicted variance $\hat{\sigma}_i^2$, enhancing robustness against noisy observations, whereas the logarithmic term introduces a regularization penalty that prevents variance inflation. Consequently, this dual head learning paradigm yields an estimate of aleatoric uncertainty representing the irreducible, data inherent unpredictability that persists despite perfect model optimization.

\subsubsection{Epistemic Uncertainty via MC Dropout}
Aleatoric uncertainty does not vanish with  more samples, whereas \emph{epistemic uncertainty} reflects model ambiguity due to limited and sparse training data and is precisely what active sensing aims to reduce. Because UAV driven measurements are expensive, many parameter configurations can explain the training set equally well, implying a non degenerate posterior over weights. Let $\mathcal{D}_{\mathrm{train}}=\{(\mathbf{X}_n,\mathbf{P}_n)\}_{n=1}^{N}$ denote the training set. Here, $\mathcal{D}_{\mathrm{train}}$ denotes the offline training dataset and should not be confused with the online cumulative measurement set $\mathcal{Y}_k$ in Sec.~\ref{subsec:models}. A Bayesian treatment seeks the weight posterior
\begin{equation}
    p(\boldsymbol{\theta} \mid \mathcal{D}_{\mathrm{train}}) = \frac{p(\mathcal{D}_{\mathrm{train}} \mid \boldsymbol{\theta})\,p(\boldsymbol{\theta})}{p(\mathcal{D}_{\mathrm{train}})},
    \label{eq:bayes_posterior}
\end{equation}
where $p(\boldsymbol{\theta})$ is a prior and $p(\mathcal{D}_{\mathrm{train}})=\int p(\mathcal{D}_{\mathrm{train}} \mid \boldsymbol{\theta})p(\boldsymbol{\theta})\,d\boldsymbol{\theta}$ is the evidence. The predictive distribution for a new input $\mathbf{X}^*$ is obtained by marginalizing over this posterior:
\begin{equation}
    p(\mathbf{P}^* \mid \mathbf{X}^*, \mathcal{D}_{\mathrm{train}}) = \int p(\mathbf{P}^* \mid \mathbf{X}^*, \boldsymbol{\theta})\, p(\boldsymbol{\theta} \mid \mathcal{D}_{\mathrm{train}})\, d\boldsymbol{\theta}.
    \label{eq:predictive}
\end{equation}

Computing Eq.~\eqref{eq:bayes_posterior} exactly is intractable for deep networks. Variational inference (VI) overcomes this by approximating $p(\boldsymbol{\theta} \mid \mathcal{D}_{\mathrm{train}})$ with a tractable variational distribution $q_{\phi}(\boldsymbol{\theta})$. This is achieved by maximizing the evidence lower bound:
\begin{equation}
    \mathcal{L}_{\mathrm{ELBO}}(\phi) = \mathbb{E}_{q_{\phi}(\boldsymbol{\theta})}\!\left[\log p(\mathcal{D}_{\mathrm{train}} \mid \boldsymbol{\theta})\right] - \mathrm{KL}\!\left(q_{\phi}(\boldsymbol{\theta})\,\|\,p(\boldsymbol{\theta})\right).
    \label{eq:kl_elbo}
\end{equation}
Maximizing $\mathcal{L}_{\mathrm{ELBO}}(\phi)$ is mathematically equivalent to minimizing the Kullback--Leibler (KL) divergence between the approximate distribution $q_{\phi}(\boldsymbol{\theta})$ and the true posterior $p(\boldsymbol{\theta} \mid \mathcal{D}_{\mathrm{train}})$. Following~\cite{Yarin2016BNN_dropout}, MC dropout provides an efficient VI approximation by interpreting dropout masks as sampling from a Bernoulli variational family. Concretely, inserting dropout (especially in intermediate decoder layers) induces stochastic network realizations that can be viewed as drawing weight samples $\hat{\boldsymbol{\theta}} \sim q_{\phi}(\boldsymbol{\theta})$. During training, the expected data fit term $\mathbb{E}_{q_{\phi}}[\log p(\mathcal{D}_{\mathrm{train}} \mid \boldsymbol{\theta})]$ is optimized via stochastic dropout samples (e.g., the heteroscedastic NLL in Eq.~\eqref{eq:nll}), while the KL regularizer $\mathrm{KL}(q_{\phi}\,\|\,p(\boldsymbol{\theta}))$ is commonly approximated by weight decay under a Gaussian prior.

Unlike standard dropout, which is disabled at test time, MC dropout keeps dropout active during inference~\cite{Yarin2016BNN_dropout}. Given the current observations $(\mathcal{Y}_k, \mathcal{M}_k)$, $S$ stochastic forward passes---each corresponding to a different sampled \emph{sub-network}---approximate the predictive integral in Eq.~\eqref{eq:predictive}, yielding an ensemble $\{\hat{P}_{k}^{(j)}, \hat{\sigma}_{k}^{2,(j)}\}_{j=1}^{S}$. The posterior mean radio map is $\bar{P}_{k} = \frac{1}{S}\sum_{j=1}^{S} \hat{P}_{k}^{(j)}$. Combining MC dropout with the heteroscedastic variance head enables a principled uncertainty decomposition. For each grid cell $i$, the predictive variance is decomposed into aleatoric and epistemic components, which can be estimated from the $S$ stochastic passes as:
\begin{align}
        \mathrm{var}(P_i^* \mid \mathbf{X}^*, \mathcal{D}_{\mathrm{train}})
        \hspace{-0.5mm}&\approx \hspace{-0.5mm}\underbrace{\mathbb{E}_{q_{\phi}(\boldsymbol{\theta})}\!\left[\hat{\sigma}_i^2(\mathbf{X}^*;\boldsymbol{\theta})\right]}_{\text{aleatoric}} \hspace{-0.5mm}+ \hspace{-0.5mm}\underbrace{\mathrm{var}_{q_{\phi}(\boldsymbol{\theta})}\!\big(\hat{P}_i(\mathbf{X}^*;\boldsymbol{\theta})\big)}_{\text{epistemic}} \nonumber \\
&\approx \frac{1}{S}\sum_{j=1}^{S} \hspace{-0.5mm}\left(\hat{\sigma}_i^{2,(j)} + (\hat{P}_i^{(j)}-\bar{P}_i)^2\right).
    \label{eq:mc_var_decomp}
\end{align}
    
For active sensing, epistemic uncertainty indicates where additional measurements are informative, whereas aleatoric uncertainty is only partially reducible. In practice, the aggregate uncertainty map used for planning therefore weights the two empirical components with a planning coefficient $\eta$:
\begin{equation}
    U_{k} = \frac{1}{S} \sum_{j=1}^{S} (\hat{P}_{k}^{(j)} - \bar{P}_{k})^2 + \eta \cdot \frac{1}{S} \sum_{j=1}^{S} \hat{\sigma}_{k}^{2,(j)}.
    \label{eq:uncertainty_total}
\end{equation}
The resulting belief state $(\bar{P}_{k}, U_{k})$ serves as the output of the perception module. High uncertainty regions indicate unreliable predictions and are thus prioritized by the downstream planner for information gathering. In Stage~II, the reinforcement learning agent aggregates $U_k$ into a scalar $\Sigma_k=\sum_{\mathbf{q}\in\mathcal{Q}}U_k(\mathbf{q})$ and is rewarded for its reduction (Eq.~\eqref{eq:reward}), providing a computable information gain surrogate when the ground-truth reconstruction error is unavailable online.

\begin{algorithm}[t]
    \caption{Training for the masked radio map reconstruction.}
    \label{alg:ssrm}
    \begin{algorithmic}[1]
        \REQUIRE Training data $\{(P, \mathcal{H})\}$; total number of training epochs $N_{\text{epoch}}$; radio sampling schedule $\{p_r^{(t)}\}_{t=1}^{N_{\text{epoch}}}$ decreasing across epochs; patch size $s$; random geometric retention $p_g \sim \mathrm{Uniform}(p_g^{\min}, p_g^{\max})$; MC samples $S$
        \ENSURE Trained Bayesian U-Net $f_\theta$; belief state $(\bar{P}_k, U_k)$ at step $k$
        \FOR{$t = 1, 2, \dots, N_{\text{epoch}}$}
        \STATE Set radio retention probability $p_r \leftarrow p_r^{(t)}$
        \FOR{each minibatch $(P,\mathcal{H})$}
        \STATE Sample sparse radio mask $\mathbf{M}_r \sim \mathrm{Bernoulli}(p_r)$ on a pixel-wise basis
        \STATE Sample $p_g$ and generate geometric mask $\mathbf{M}_g$ by retaining $s{\times}s$ patches w.p. $p_g$
        \STATE Form sparse radio observation $\mathbf{P}^{\mathrm{obs}} \leftarrow P \odot \mathbf{M}_r$
        \STATE Form masked input $\mathbf{X} \leftarrow \text{Concat}(\mathbf{P}^{\mathrm{obs}}, \mathcal{H} \odot \mathbf{M}_g)$
        \STATE Forward pass with dropout: $f_\theta(\mathbf{X}) \rightarrow (\hat{P}, \hat{\sigma}^2)$
        \STATE Update $\theta$ by minimizing Eq.~\eqref{eq:nll}
        \ENDFOR
        \ENDFOR
        \STATE {Online inference at step $k$:} rasterize $(\mathcal{Y}_k, \mathcal{M}_k)$ into $\mathbf{X}_k$ via Eq.~\eqref{eq:input_concat}
        \FOR{$j = 1, 2, \dots, S$}
        \STATE Stochastic pass: $f_\theta(\mathbf{X}_k) \rightarrow (\hat{P}_{k}^{(j)}, \hat{\sigma}_{k}^{2,(j)})$
        \ENDFOR
        \STATE Compute $\bar{P}_k$ and $U_k$ using Eq~\eqref{eq:uncertainty_total}; return $(\bar{P}_k, U_k)$
    \end{algorithmic}
\end{algorithm}

Algorithm~\ref{alg:ssrm} summarizes Stage~I, encompassing offline training with dual masked inpainting and online inference via MC dropout. The key output is the frozen predictor $f_{\theta^\star}$, which produces a predictive belief state $(\bar{P}_k, U_k)$ for observation patterns encountered during deployment.

\begin{figure*}[t]
    \centering
    \includegraphics[width=0.9\linewidth]{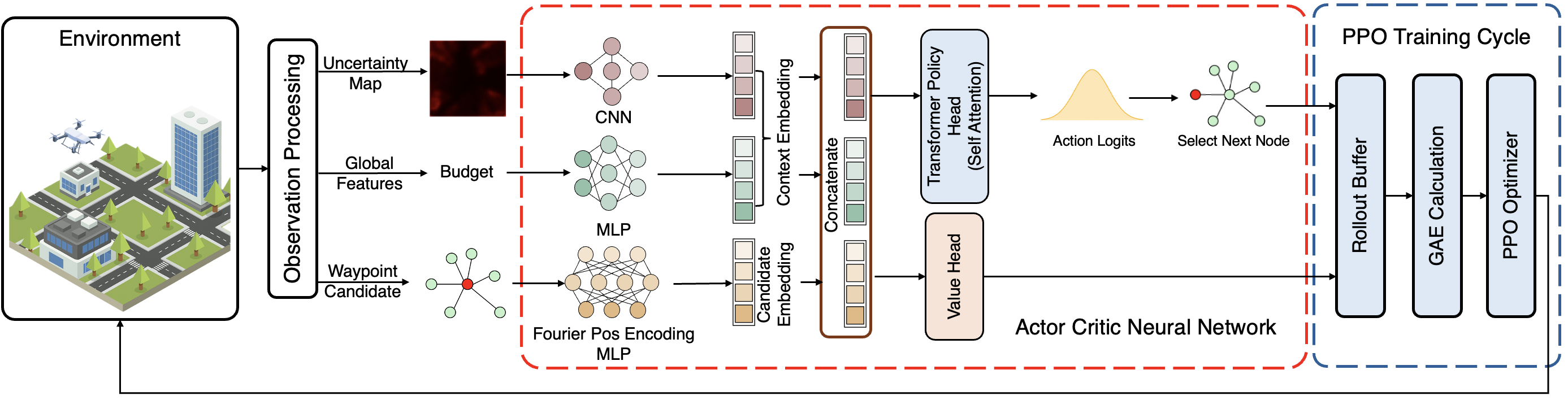}
    \caption{Transformer actor--critic and PPO loop for waypoint selection.}
    \label{fig:rl_framework}
\end{figure*}

\section{Stage II: Budget Constrained Active Exploration via DRL}\label{sec:drl_method}
While Stage~I successfully yields a Bayesian reconstructor capable of quantifying prediction uncertainty, it relies on static observational patterns and does not dictate how the UAV should continually acquire future measurements. To bridge this gap, Stage~II leverages the generated dense uncertainty map as an intrinsic reward surrogate to learn an active sensing policy. 

\subsection{Construction of Dynamic Local Graph}\label{subsec:dynamic_local_graph}
The operating environment $\Omega$ is continuous and initially unknown, revealed only incrementally through onboard perception. Classical planners that assume full observability and a static search space are therefore inapplicable. To bridge the gap between the continuous physical domain and tractable sequential decision making, the proposed framework adopts a dynamic graph-based abstraction and constructs a transient local Probabilistic Roadmap $\mathcal{G}_k$ online from the current perception at each decision step $k$. Unlike conventional PRMs that rely on a precomputed global topology, this dynamic PRM discretizes the currently known free space into a finite set of collision free candidate waypoints. As new observations arrive from onboard camera, the graph is regenerated to remain synchronized with the evolving occupancy map $\mathcal{M}_k$ and the radio map belief state.
Such a graph-based abstraction converts an intractable continuous 3D control problem into a categorical selection over a fixed size candidate set, guarantees collision freedom at the moment of action generation, and allows the travel budget to be accumulated exactly along the vertical and horizontal legs of each transition. Waypoints are moreover the native mission format of commercial autopilot stacks~\cite{DJI2024Waypoint} and conform to the pre-declared flight geometry expected by civil aviation frameworks~\cite{FAA2025Part108}.

At step $k$, let $\mathbf{p}_k$ denote the UAV's current position. A star topology graph $\mathcal{G}_k = (\mathcal{V}_k, \mathcal{E}_k)$ centered at $\mathbf{p}_k$ is constructed. The node set $\mathcal{V}_k$ consists of the current position and a fixed size set of $m_k$ candidate waypoints sampled from the local observation space.
Candidate waypoints are sampled uniformly within a kinematic radius $R$ centered at $\mathbf{p}_k$ and filtered by the two-leg collision constraints induced by the current occupancy map $\mathcal{M}_k$ and by the remaining budget. The node set is defined as:
\begin{equation}
    \begin{aligned}
        \mathcal{V}_k & = \{ \mathbf{p}_k \} \cup \mathcal{W}_k, \\
        \mathcal{W}_k & = \bigl\{ \mathbf{w}_{k,i} \mid i=1,\dots,m_k, \\
        &\quad \| \mathbf{w}_{k,i} - \mathbf{p}_k \|_2 \leq R,\;
        c(\mathbf{p}_k,\mathbf{w}_{k,i})\le b_k, \\
        &\quad \Gamma(\mathbf{p}_k,\mathbf{w}_{k,i})\ \text{is collision free}\bigr\}.
    \end{aligned}
    \label{eq:graph_nodes}
\end{equation}
The edge set $\mathcal{E}_k$ comprises directed edges from $\mathbf{p}_k$ to each valid candidate $\mathbf{w}_{k,i} \in \mathcal{W}_k$, with each edge representing the corresponding route $\Gamma(\mathbf{p}_k,\mathbf{w}_{k,i})$.

Beyond geometric feasibility, each node is augmented with radio map attributes derived directly from the probabilistic reconstruction.
At step $k$, the Bayesian U-Net produces a posterior mean radio map $\bar{P}_k$ and a corresponding uncertainty map $U_k$, consistent with Eq.~\eqref{eq:uncertainty_total}.
For each candidate waypoint $\mathbf{w}_{k,i}$, the U-Net prediction is queried at its spatial location, and the resulting mean and uncertainty are concatenated with the waypoint's absolute coordinate:
\begin{equation}
    \mathbf{x}_i =
    \left[
        \mathbf{w}_{k,i} ,\;
        \bar{P}_k(\mathbf{w}_{k,i}),\;
        U_k(\mathbf{w}_{k,i})
        \right].
    \label{eq:node_features}
\end{equation}
This parameterization couples geometric reachability with predicted radio informativeness, allowing the policy to favor waypoints that are expected to reduce reconstruction uncertainty.

\subsection{Markov Decision Process (MDP) Formulation}
The dynamic graph formulation naturally admits a Markov Decision Process (MDP) abstraction. In active mapping, directly planning over the ever-expanding trajectory history violates the Markov property. To resolve this, the Bayesian U-Net compresses the cumulative observations ($\mathcal{Y}_k, \mathcal{M}_k$) into the belief state $(\bar{P}_k, U_k)$, while the Dynamic PRM distills the local free space into the candidate set $\mathcal{W}_k$. Together, they form a sufficient statistic that summarizes all task-relevant past information. Conditioned on this aggregated state, future transitions are independent of the specific historical path taken, thereby satisfying the Markov property.

\subsubsection{State Space}
The state $s_k \in \mathcal{S}$ encodes both the global belief and the local candidate set at decision step $k$:
\begin{equation}
    s_k = \left( \mathbf{s}^{\mathrm{global}}_k,\; \{ \mathbf{x}_i \}_{i=1}^{m_k} \right),
\end{equation}
where $\mathbf{s}^{\mathrm{global}}_k$ denotes the global context, consisting of the dense uncertainty map $U_k \in \mathbb{R}^{H \times W}$ and a low dimensional agent state vector that concatenates the normalized remaining budget $b_k$ with the current UAV position $\mathbf{p}_k$. The set $\{ \mathbf{x}_i \}_{i=1}^{m_k}$ is a fixed size collection of candidate waypoint attributes extracted from $\mathcal{G}_k$, as defined in Eq.~\eqref{eq:node_features}.

\subsubsection{Action Space}
The action space $\mathcal{A}(s_k)$ corresponds to the set of outgoing edges in $\mathcal{G}_k$. An action $a_k \in \{1, \dots, m_k\}$ selects a candidate waypoint index; specifically, $a_k=i$ commands the UAV to first adjust its altitude vertically and then fly horizontally from $\mathbf{p}_k$ to $\mathbf{w}_{k,i}$ along $\Gamma(\mathbf{p}_k,\mathbf{w}_{k,i})$. Since $\mathcal{W}_k$ is regenerated online from local perception, the action set evolves over time in spatial configuration, while its cardinality $m_k$ remains fixed.

\subsubsection{Reward Structure}
The reward $r_k$ is designed to maximize uncertainty reduction under a finite budget. The aggregate map uncertainty at step $k$ is defined as
$\Sigma_k = \sum_{\mathbf{q} \in \mathcal{Q}} U_k(\mathbf{q})$, where $\mathcal{Q}$ denotes the set of grid locations. The instantaneous reward is then given by
\begin{equation}
    r_k =
    \begin{cases}
        \displaystyle
        \alpha \cdot \frac{\Sigma_{k-1} - \Sigma_{k}}{\Sigma_{k-1}}, & \text{if } b_{k+1} > 0,   \\[10pt]
        \displaystyle
        -\beta \cdot \Sigma_{k},                                     & \text{if } b_{k+1} \le 0.
    \end{cases}
    \label{eq:reward}
\end{equation}
For intermediate steps where $b_{k+1}>0$, the agent is rewarded proportionally to the normalized reduction in global uncertainty. Upon budget exhaustion where $b_{k+1}\le 0$, the reward switches to the terminal penalty $-\beta \Sigma_k$, which penalizes residual uncertainty at termination and aligns the learning objective with minimizing the final reconstruction uncertainty.

\subsection{Neural Policy Architecture and Optimization}
At each decision step $k$, the policy must fuse a dense global belief with a sparse, permutation invariant set of candidate waypoints. This setting poses two main challenges: (i) jointly processing high dimensional raster inputs and low dimensional proprioception, and (ii) selecting an action from an unordered candidate set of fixed cardinality $m_k$. These challenges are addressed with a hybrid architecture consisting of a dual stream global context encoder, a Fourier enhanced candidate embedding module, and a Transformer-based decision head, as illustrated in Fig.~\ref{fig:rl_framework}.
The $m_k$ candidates are treated as an unordered token set, which enables permutation invariant reasoning over feasible navigation options while conditioning on the global belief. The actor produces a categorical distribution over candidate indices, and the critic provides value estimates used to compute advantages during PPO updates.

Concretely, the policy network proceeds in three stages. First, the global context $\mathbf{s}^{\mathrm{global}}_k$ is encoded by two parallel streams. The dense uncertainty map $U_k \in \mathbb{R}^{H \times W}$ is encoded by a 3-layer convolutional neural network (CNN) with $3 \times 3$ kernels and channel dimensions $\{32,64,64\}$, using ReLU activations and max-pooling to produce a compact spatial feature. In parallel, a lightweight multi-layer perceptron (MLP) embeds a low-dimensional proprioceptive vector formed by concatenating the normalized remaining budget $b_k$ with the current UAV position $\mathbf{p}_k$. The two features are concatenated and projected to a unified context embedding $\mathbf{z}_{\mathrm{ctx}} \in \mathbb{R}^{d_{\mathrm{embed}}}$ that conditions subsequent candidate evaluation.

Given this context embedding, each candidate waypoint is represented by explicitly encoding both geometry and local radio attributes. Since standard MLPs can exhibit spectral bias and underfit high-frequency spatial variations, {Fourier feature mapping} is employed. For a coordinate vector $\mathbf{u} \in \mathbb{R}^3$, the mapping $\xi(\cdot)$ lifts the input to a multi-frequency domain, applied element-wise:
\begin{equation}
    \xi(\mathbf{u})_{\ell} =
    \begin{cases}
        \sin\!\big(2^{(\ell-1)/2}\pi \mathbf{u}\big), & \ell~is~odd\\
        \cos\!\big(2^{(\ell-2)/2}\pi \mathbf{u}\big), & \ell~is~even.
    \end{cases}
\end{equation}
where $L$ is the number of frequency bands. The mapping $\xi(\cdot)$ is applied to both the absolute waypoint coordinate $\mathbf{w}_{k,i}$ and the relative displacement $\mathbf{w}_{k,i}-\mathbf{p}_k$. These geometric embeddings are concatenated with the local waypoint attributes in $\mathbf{x}_i$ defined in Eq.~\eqref{eq:node_features} and passed through a fusion MLP, yielding the candidate embedding $\mathbf{e}_i$.

Finally, to reason jointly over the candidate set and avoid redundant sampling among spatially clustered options, a Transformer-based architecture is employed to model interdependencies among candidates while remaining permutation invariant. To condition local candidate features on the current budget and map state, the global context embedding is injected into each candidate token:
\begin{equation}
    \tilde{\mathbf{e}}_i = \mathbf{e}_i + \mathbf{W}_{\mathrm{ctx}} \mathbf{z}_{\mathrm{ctx}},
\end{equation}
where $\mathbf{W}_{\mathrm{ctx}}$ is a learned linear projection. The conditioned tokens $\{ \tilde{\mathbf{e}}_i \}_{i=1}^{m_k}$ are then processed by a standard Transformer encoder with multihead self attention:
\begin{equation}
    \text{Attention}(\mathbf{Q}, \mathbf{K}, \mathbf{V}) = \text{softmax}\left(\frac{\mathbf{Q}\mathbf{K}^T}{\sqrt{d_{\text{key}}}}\right)\mathbf{V}.
\end{equation}
where $d_{\text{key}}$ denotes the key dimensionality. This design is permutation invariant and operates over the fixed size candidate set of cardinality $m_k$. Finally, a shared decision head maps the refined tokens to scalar logits, and a Softmax defines the policy distribution $\pi_{\vartheta}(a_k|s_k)$.

To train this actor critic, the policy $\pi_\vartheta$ and value function $V_\omega$ are jointly optimized using PPO~\cite{Schulman2017PPO}, which maximizes a clipped surrogate objective together with a value regression term and an entropy regularizer:
\begin{equation}
    \mathcal{L}(\vartheta,\omega) = \hat{\mathbb{E}}_k \left[ \mathcal{L}^{\text{CLIP}}_k(\vartheta) - c_1 \mathcal{L}^{\text{VF}}_k(\omega) + c_2 \mathcal{S}[\pi_\vartheta](s_k) \right],
    \label{eq:total_loss}
\end{equation}
where $c_1$ and $c_2$ are coefficients balancing value estimation accuracy and exploration, respectively.

The main policy update term is the clipped surrogate objective. Let $\rho_k(\vartheta) = \frac{\pi_\vartheta(a_k|s_k)}{\pi_{\vartheta_{\mathrm{old}}}(a_k|s_k)}$ denote the probability ratio between the current and old policies. The clipped policy loss $\mathcal{L}^{\text{CLIP}}_k$ encourages probability increases for advantageous actions while bounding the update magnitude:
\begin{equation}
    \mathcal{L}^{\text{CLIP}}_k = \min \left( \rho_k(\vartheta) \hat{A}_k, \; \operatorname{clip}(\rho_k(\vartheta), 1-\epsilon, 1+\epsilon) \hat{A}_k \right),
\end{equation}
where $\epsilon$ is a hyperparameter controlling the clipping range, e.g., $\epsilon=0.2$, and $\hat{A}_k$ is the estimated advantage. The clipping mechanism ensures that $\rho_k(\vartheta)$ does not deviate significantly from unity, thereby maintaining training stability.

To compute $\hat{A}_k$ in Eq.~\eqref{eq:total_loss}, generalized advantage estimation (GAE) is employed to reduce gradient variance while maintaining manageable bias. The temporal difference (TD) error $\delta_k$ is defined as:
\begin{equation}
    \delta_k = r_k + \gamma V_\omega(s_{k+1}) - V_\omega(s_k),
\end{equation}
where $V_\omega$ is the critic. For a rollout of length $T_{\mathrm{roll}}$, the advantage estimate is computed as an exponentially weighted sum of TD errors:
\begin{equation}
    \hat{A}_k = \sum_{l=0}^{T_{\mathrm{roll}}-k-1} (\gamma \lambda)^l \delta_{k+l},
\end{equation}
where $\gamma$ is the discount factor and $\lambda$ is the GAE smoothing parameter.
Algorithm~\ref{alg:ppo_planner} summarizes the PPO training procedure for the proposed graph based waypoint planner.
\begin{algorithm}[t]
    \caption{PPO training for the waypoint selection policy.}
    \label{alg:ppo_planner}
    \begin{algorithmic}[1]
        \REQUIRE Simulator environment; candidate generator yielding $\mathcal{W}_k=\{\mathbf{w}_{k,i}\}_{i=1}^{m_k}$; node attributes in Eq.~\eqref{eq:node_features}; policy $\pi_\vartheta$; critic $V_\omega$; PPO hyperparameters $(\gamma,\lambda,\epsilon,c_1,c_2)$; rollout length $T_{\mathrm{roll}}$; update epochs $N_{\mathrm{epoch}}$
        \ENSURE Trained planner parameters $(\vartheta,\omega)$
        \FOR{iteration $=1,2,\dots$}
        \STATE Set $\vartheta_{\mathrm{old}} \leftarrow \vartheta$ and collect a rollout:
        \FOR{$k=0,1,\dots,T_{\mathrm{roll}}-1$}
        \STATE Observe $s_k=(\mathbf{s}^{\mathrm{global}}_k,\{\mathbf{x}_i\}_{i=1}^{m_k})$ and sample $a_k \sim \pi_{\vartheta_{\mathrm{old}}}(\cdot \mid s_k)$
        \STATE Execute $a_k$; receive $r_k$ and $s_{k+1}$; store $(s_k,a_k,r_k,\log \pi_{\vartheta_{\mathrm{old}}}(a_k|s_k))$
        \ENDFOR
        \STATE Compute advantages $\{\hat{A}_k\}$ with GAE and returns $\hat{R}_k=\hat{A}_k+V_\omega(s_k)$; normalize $\{\hat{A}_k\}$ within the batch
        \FOR{epoch $=1,2,\dots,N_{\mathrm{epoch}}$}
        \STATE Update $(\vartheta,\omega)$ by optimizing Eq.~\eqref{eq:total_loss} with Adam using minibatches; apply entropy regularization and gradient clipping
        \ENDFOR
        \ENDFOR
    \end{algorithmic}
\end{algorithm}

\subsection{Iterative Perception--Action Loop}
During online execution, 3D-URAM operates as a closed-loop perception--action system, summarized in Algorithm~\ref{alg:online_deploy}. At each decision step $k$, the UAV updates its spatial and electromagnetic awareness based on the cumulative radio measurements $\mathcal{Y}_k$ and the dynamically expanded occupancy map $\mathcal{M}_k$.

\begin{algorithm}[t]
    \caption{Online deployment of 3D-URAM.}
    \label{alg:online_deploy}
    \begin{algorithmic}[1]
        \REQUIRE Frozen Bayesian U-Net $f_{\theta^\star}$; learned policy $\pi$; initial position $\mathbf{p}_0$; initial maps $(\mathcal{Y}_0,\mathcal{M}_0)$; travel budget $B$; MC samples $S$
        \ENSURE Trajectory $\mathcal{T}_\pi$ and measurements $\mathcal{Y}_T$
        \STATE Initialize $k \leftarrow 0$ and remaining budget $b_0 \leftarrow B$
        \WHILE{$b_k > 0$}
        \STATE Belief update: $(\bar{P}_k, U_k) \leftarrow f_{\theta^\star}(\mathcal{Y}_k,\mathcal{M}_k)$ using $S$ stochastic forward passes
        \STATE Construct a transient local graph $\mathcal{G}_k$ and budget-feasible candidates $\mathcal{W}_k$ from $(\mathcal{M}_k,b_k)$ (Sec.~\ref{subsec:dynamic_local_graph})
        \STATE Form node attributes $\{\mathbf{x}_i\}_{i=1}^{m_k}$ via Eq.~\eqref{eq:node_features}
        \STATE Select $a_k \sim \pi(\cdot \mid s_k)$ and set $\mathbf{p}_{k+1} \leftarrow \mathbf{w}_{k,a_k}$
        \STATE Execute $\Gamma(\mathbf{p}_k,\mathbf{p}_{k+1})$ (vertical first, then horizontal); collect new measurements $\Psi_k$
        \STATE Update $\mathcal{Y}_{k+1} \leftarrow \mathcal{Y}_k \cup \Psi_k$ and expand $\mathcal{M}_{k+1}$ from onboard perception
        \STATE Update budget $b_{k+1} \leftarrow b_k-c(\mathbf{p}_k,\mathbf{p}_{k+1})$; set $k \leftarrow k+1$
        \ENDWHILE
    \end{algorithmic}
\end{algorithm}

Specifically, the Bayesian network maps past observations into a dense belief state $(\bar{P}_k, U_k)$, which is then projected onto the dynamically constructed valid candidate waypoints $\mathcal{W}_k$. By extracting the attributes for each candidate as in Eq.~\eqref{eq:node_features}, the continuous belief is discretized into a format directly consumable by the Transformer policy. The policy then selects an optimal waypoint $\mathbf{p}_{k+1}$ to maximize the expected information gain. Navigating toward this waypoint yields new measurements $\Psi_k$ and expands the perceived environment $\mathcal{M}_{k+1}$, driving the next iteration. This active sensing loop repeats until the UAV exhausts its travel budget $B$.

\begin{figure*}[t]
    \centering
    \includegraphics[width=1.0\linewidth]{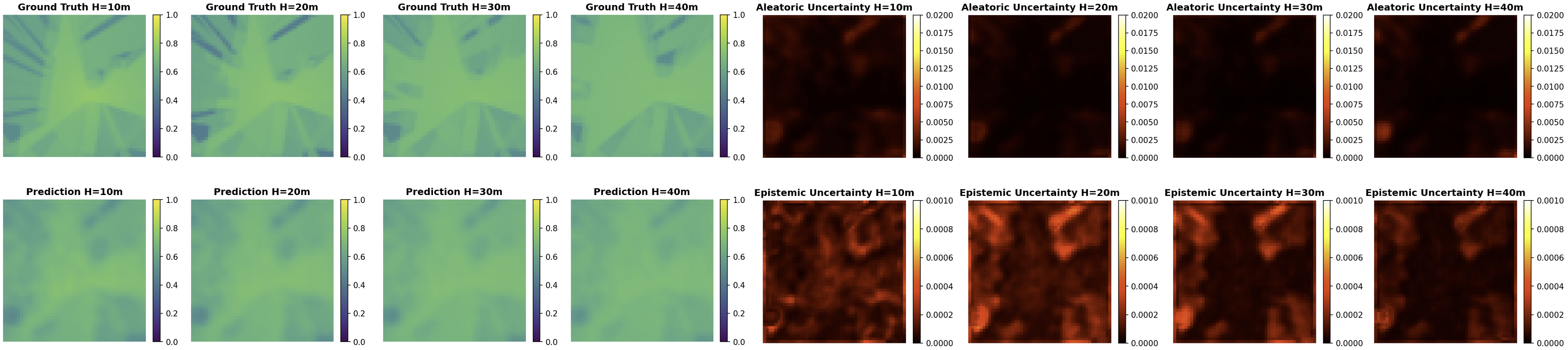}
    \caption{Stage-I predictions and decomposed uncertainty across four altitude layers ($h=10$, $20$, $30$, and $40~\mathrm{m}$). The left block shows the ground-truth radio maps (top) and Bayesian U-Net predictions (bottom); the right block shows aleatoric uncertainty (top) and epistemic uncertainty (bottom).}
    \label{fig:uncertainty_vis}
\end{figure*}

\subsection{Computational Complexity}
To ensure real-time feasibility for onboard deployment, the per step computational complexity of the replanning loop is analyzed. Let $H \times W$ denote the raster resolution of each altitude layer, $Z$ the number of altitude layers, $S$ the number of stochastic forward passes, and $m_k=|\mathcal{W}_k|$ the fixed number of candidate waypoints. Voxelization and map fusion scale as $\mathcal{O}(ZHW)$, while dynamic PRM sampling and collision checking scale as $\mathcal{O}(m_k)$. Updating the Bayesian belief over all altitude layers requires $S$ forward passes and scales as $\mathcal{O}(SZHW)$; encoding the resulting uncertainty volume with the CNN incurs $\mathcal{O}(ZHW)$, and Transformer self attention over the candidate set incurs $\mathcal{O}(m_k^2)$ per layer. Consequently, absorbing the lower-order terms, the overall per-step complexity is
\begin{equation}
    \mathcal{O}\!\left(SZHW + m_k^2\right).
    \label{eq:step_complexity}
\end{equation}
Since $Z$, $S$, and $m_k$ are fixed small constants in practice, the replanning overhead scales linearly with the environmental map resolution.

This bounded complexity presents a substantial advantage over classical probabilistic methods. For instance, dense GP inference incurs an $\mathcal{O}(N_k^3 + N_k Z H W)$ complexity, where $N_k \triangleq |\mathcal{Y}_k|$ denotes the number of accumulated samples. As $N_k$ monotonically increases during exploration, the cubic complexity rapidly becomes a bottleneck for onboard replanning. While deterministic deep models, such as Spectrum Surveying, achieve a lower complexity of $\mathcal{O}(Z H W)$ using a single forward pass, they fail to capture the epistemic uncertainty essential for directing active exploration toward genuinely unexplored regions.
By combining aleatoric uncertainty learned via a heteroscedastic NLL loss, a lightweight MC dropout ensemble ($S{=}10$) for epistemic uncertainty, and a fixed cardinality candidate set via the dynamic PRM, 3D-URAM delivers uncertainty estimates for exploration without suffering the severe scalability issues of GP-based methods.

\section{Experiments}
\subsection{Dataset Generation}\label{sec:dataset_gen}
To train and evaluate 3D-URAM, we conduct a large scale radio map dataset using NVIDIA Sionna, a GPU accelerated physics based simulator with deterministic ray tracing that models reflection, diffraction, and scattering. The ray tracing dataset is generated via the following pipeline:
\begin{enumerate}
    \item \textbf{Scene extraction:} 150 urban scenes are extracted from OpenStreetMap spanning diverse building layouts and densities. Each scene covers a $200 \times 200~\mathrm{m}^2$ area.
    \item \textbf{Geometry and materials:} Buildings are approximated as cuboids, and electromagnetic material models are assigned to the ground, walls, and rooftops, as listed in Table~\ref{tab:sim_params}, to capture LoS blockage and urban canyon propagation effects.
    \item \textbf{Transceiver setup:} 25 transmitters are randomly deployed per scene, and receiver locations are placed on uniform grids across four altitude layers ($h \in \{10, 20, 30, 40\}$~m).
    \item \textbf{Ray tracing and slicing:} The Sionna~\cite{Jakob2023Sionna} is used to compute path loss, yielding $150 \times 25 \times 4 = 15{,}000$ radio map slices.
    \item \textbf{Preprocessing:} Path loss maps are clipped to $[0, 150]$~dB to suppress negligible weak signals and mitigate extreme near field singularities, then linearly mapped to $[0, 255]$ and quantized to 8-bit unsigned integers, denoted \texttt{uint8}, producing grayscale images $\mathbf{P} \in \{0, \dots, 255\}^{H \times W}$.
\end{enumerate}

\begin{table}[t]
    \centering
    \caption{Ray Tracing Simulation Parameters.}
    \label{tab:sim_params}
    \renewcommand{\arraystretch}{1.2}
    \begin{tabular}{lc}
        \toprule
        {Parameter}             & {Value}                \\
        \midrule
        Carrier Frequency       & 2.4 GHz                \\
        Transmit Power          & 0 dBm                  \\
        Antenna Pattern         & Isotropic              \\
        Diffraction             & Enabled                \\
        Refraction              & Disabled               \\
        Rooftop Material        & ITU\_brick             \\
        Wall Material           & ITU\_concrete          \\
        Ground Material         & ITU\_very\_dry\_ground \\
        \bottomrule
    \end{tabular}
\end{table}

Finally, the dataset is partitioned by environment using a 7:3 train/test split, with 105 training scenes and 45 test scenes. This strict separation ensures that evaluation is performed on entirely unseen building layouts, thereby providing a stringent assessment of generalization across diverse urban geometries.

\subsection{Uncertainty Network}
The radio map prediction network is trained on the ray tracing dataset described in Sec.~\ref{sec:dataset_gen}.
During training, a curriculum learning schedule gradually increases the input masking ratio from $0.8$ to $0.98$ to ensure robust convergence. At inference, MC dropout remains active to estimate epistemic uncertainty. Table~\ref{tab:train_params} summarizes the training hyperparameters.
We compare three planning baselines: covariance matrix adaptation evolution strategy (CMA-ES), Monte Carlo Tree Search (MCTS), and random sampling.

\begin{table*}[h]
    \centering
    \small
    \caption{RMSE for Budgets, Planners, and Reconstructors.}
    \label{tab:baseline_matrix}
    \begin{tabular}{llccc ccc ccc}
        \toprule
        \multirow{2}{*}{{Budget}}          &
        \multirow{2}{*}{{Planning Method}} &
        \multicolumn{3}{c}{{GP-DKL}}       &
        \multicolumn{3}{c}{{Spectrum Surveying}}         &
        \multicolumn{3}{c}{\textbf{Bayesian U-Net [Ours]}}                                                                             \\
        \cmidrule(lr){3-5} \cmidrule(lr){6-8} \cmidrule(lr){9-11}
                                           &          & Best   & Avg    & Worst  & Best   & Avg    & Worst  & Best   & Avg    & Worst  \\
        \midrule
        \multirow{4}{*}{$B=1$}
                                           & Proposed & 0.5034 & 0.5187 & 0.5254 & 0.1411 & 0.1522 & 0.1648 & \fcolorbox{red}{white}{0.0519} & \fcolorbox{red}{white}{0.0536} & \fcolorbox{red}{white}{0.0548} \\
                                           & CMA-ES   & 0.5096 & 0.5173 & 0.5325 & 0.1520 & 0.1742 & 0.1898 & 0.0663 & 0.0694 & 0.0721 \\
                                           & MCTS     & 0.5084 & 0.5113 & 0.5237 & 0.1499 & 0.1635 & 0.1840 & 0.0592 & 0.0623 & 0.0677 \\
                                           & Random   & 0.4933 & 0.5538 & 0.6002 & 0.1549 & 0.2194 & 0.2688 & 0.0603 & 0.0714 & 0.0799 \\
        \cmidrule{2-11}
        \multirow{4}{*}{$B=2$}
                                           & Proposed & 0.4643 & 0.4791 & 0.4911 & 0.1103 & 0.1218 & 0.1402 & \fcolorbox{red}{white}{0.0435} & \fcolorbox{red}{white}{0.0452} & \fcolorbox{red}{white}{0.0479} \\
                                           & CMA-ES   & 0.4770 & 0.4926 & 0.5067 & 0.1247 & 0.1321 & 0.1521 & 0.0495 & 0.0536 & 0.0588 \\
                                           & MCTS     & 0.4792 & 0.4866 & 0.5148 & 0.1194 & 0.1333 & 0.1640 & 0.0476 & 0.0524 & 0.0553 \\
                                           & Random   & 0.4564 & 0.5687 & 0.6493 & 0.1362 & 0.1782 & 0.2531 & 0.0519 & 0.0651 & 0.0780 \\
        \cmidrule{2-11}
        \multirow{4}{*}{$B=4$}
                                           & Proposed & 0.3909 & 0.4217 & 0.4683 & 0.0664 & 0.0813 & 0.1027 & \fcolorbox{red}{white}{0.0320} & \fcolorbox{red}{white}{0.0355} & \fcolorbox{red}{white}{0.0382} \\
                                           & CMA-ES   & 0.4311 & 0.4658 & 0.4926 & 0.0791 & 0.0945 & 0.1162 & 0.0419 & 0.0487 & 0.0545 \\
                                           & MCTS     & 0.4122 & 0.4442 & 0.5029 & 0.0748 & 0.0910 & 0.1194 & 0.0416 & 0.0506 & 0.0574 \\
                                           & Random   & 0.4126 & 0.5388 & 0.5839 & 0.0667 & 0.1325 & 0.2080 & 0.0362 & 0.0581 & 0.0743 \\
        \bottomrule
    \end{tabular}
\end{table*}

\begin{table}[htbp]
    \centering
    \caption{Training Hyperparameters.}
    \label{tab:train_params}
    \renewcommand{\arraystretch}{1.2}
    \begin{tabular}{lc}
        \toprule
        {Parameter}               & {Value}                     \\
        \midrule
        \multicolumn{2}{c}{\textbf{Network Training}}           \\
        \midrule
        Input Resolution          & $50 \times 50$              \\
        Optimizer                 & Adam                        \\
        Initial Learning Rate     & $1 \times 10^{-3}$          \\
        Learning Rate Decay       & Cosine Annealing            \\
        Batch Size                & 64                          \\
        Dropout Rate              & 0.1                         \\
        Masking Ratio Strategy    & Linear from $0.8$ to $0.98$ \\
        \midrule
        \multicolumn{2}{c}{\textbf{RL Policy Training}}         \\
        \midrule
        Optimizer                 & Adam                        \\
        Learning Rate             & $3 \times 10^{-4}$          \\
        Discount Factor $\gamma$  & 0.99                        \\
        GAE Parameter $\lambda$   & 0.97                        \\
        Clipping Range $\epsilon$ & 0.2                         \\
        Value Loss Coeff. $c_1$   & 0.5                         \\
        Entropy Coeff. $c_2$      & 0.01                        \\
        Max Gradient Norm         & 0.5                         \\
        Mini-Batch Size           & 128                         \\
        Update Epochs             & 4                           \\
        Rollout Horizon, steps    & 1024                        \\
        MC Dropout Samples $S$    & 10                          \\
        Planning Coeff. $\eta$    & 1                           \\
        \bottomrule
    \end{tabular}
\end{table}

Fig.~\ref{fig:uncertainty_vis} visualizes representative predictions together with decomposed uncertainty estimates across altitudes. The uncertainty maps exhibit clear spatial structure: high uncertainty concentrates near complex geometric boundaries and non-line-of-sight regions where shadowing, diffraction, and strong multipath effects dominate. Such low confidence regions provide a reliability cue for downstream robust UAV path planning.

The proposed Bayesian U-Net is benchmarked against three representative baselines: Spectrum Surveying~\cite{Shrestha2023Spectrumsurveying}, an autoencoder based method; GP-DKL~\cite{Wilson2016dkl}, a deep kernel learning Gaussian process; and GP-ATT, an attention based Gaussian process baseline. Two evaluation settings are considered to reflect different deployment phases:
\begin{enumerate}
    \item {Full environmental knowledge:} the complete building layout is available as a prior.
    \item {Partial observability:} only a subset of building geometries is available, representing incomplete environmental context during online exploration.
\end{enumerate}

\begin{figure}[htbp]
    \centering
    \includegraphics[width=\linewidth]{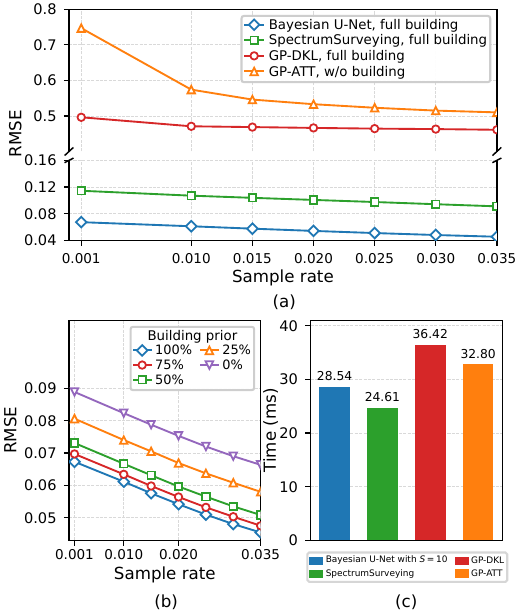}
    \caption{Proposed Bayesian U-Net comparison with Spectrum Surveying, GP-DKL, and GP-ATT.}
    \label{fig:comparison_vis}
\end{figure}

As shown in Fig.~\ref{fig:comparison_vis}(a), we compare the four methods with the full building layout. Since GP-ATT cannot fuse building information, it is evaluated without building geometries (w/o building). The proposed Bayesian U-Net outperforms all other methods. Spectrum Surveying improves upon GP-DKL but remains inferior to the Bayesian U-Net. This is because its plain autoencoder architecture lacks skip connections, resulting in the loss of fine-grained spatial details.

As shown in Fig.~\ref{fig:comparison_vis}(b), which compares the proposed Bayesian U-Net under different building priors, it can be seen that the reconstruction performance consistently improves as more building observations become available. This demonstrates that simultaneously collecting environmental priors to assist in radio map generation during pure radio signal sampling is necessary. Furthermore, as shown in Fig.~\ref{fig:comparison_vis}(c), we evaluate the inference speed of each method. Spectrum Surveying is computationally the fastest at 24.61 ms due to its single deterministic forward pass. The proposed Bayesian U-Net, despite $S=10$ stochastic passes for uncertainty quantification, achieves a highly competitive inference time of 28.54 ms. Both cases are well within real-time limits and significantly faster than the 36.42 ms required by GP-DKL.

\subsection{RL Path Planning}
\begin{figure}[t]
    \centering
    \includegraphics[width=0.8\linewidth]{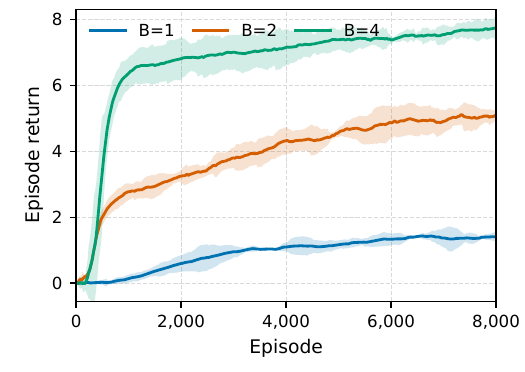}
    \caption{PPO training convergence under the three travel budgets. Larger budgets exhibit greater variability in episode returns.}
    \label{fig:rl_convergence}
\end{figure}

Following the supervised learning phase, the reinforcement learning agent is trained for active sensing. A key objective of 3D-URAM is to navigate and reconstruct radio maps in environments that are unseen by the underlying prediction model.
To strictly enforce this generalization requirement, the training set for the RL agent is constructed exclusively from the {hold out test set} of the uncertainty network, as defined in Sec.~\ref{sec:dataset_gen}. Specifically, {10 representative urban environments} are selected from the 45 test scenes. Within each selected environment, {10 distinct transmitter configurations} are utilized, forming a focused dataset for policy optimization. This setup ensures that the RL agent cannot rely on memorized environmental features from the prediction network's training phase. As shown in Fig.~\ref{fig:rl_convergence}, episode returns rise rapidly within the first 1000 episodes and stabilize over 8000 training episodes under all three budgets. Larger budgets yield higher final returns by allowing greater cumulative uncertainty reduction. Larger budgets also lead to greater variability in episode returns, which may arise from the accumulated effects of stochastic action selection and differing observation sequences over longer exploration horizons.

To comprehensively evaluate the versatility and robustness of the proposed exploration policy, its cooperative performance when coupled with varying reconstruction back ends is examined. Specifically, the agent trained via the proposed framework is integrated with three representative predictors: the kernel based {GP-DKL}~\cite{Wilson2016dkl}, the deep learning based {Spectrum Surveying}~\cite{Shrestha2023Spectrumsurveying}, and the proposed Bayesian U-Net.
To benchmark the planning efficiency against established methods, the proposed approach is compared with three alternative strategies: the evolutionary algorithm CMA-ES~\cite{Nikolaus2016CMA}, the heuristic search based MCTS~\cite{Maciej2021MCTS}, and random sampling approach. The comparative results are presented in Table~\ref{tab:baseline_matrix}.

As detailed in Table~\ref{tab:baseline_matrix}, the proposed Bayesian U-Net consistently demonstrates superior reconstruction fidelity across varying trajectory length budgets $B$. In particular, when paired with the proposed RL policy, the Bayesian U-Net achieves the lowest RMSE in all tested scenarios, with average RMSE decreasing from $0.0536$ at $B{=}1\,\mathrm{km}$ to $0.0452$ at $B{=}2\,\mathrm{km}$ and $0.0355$ at $B{=}4\,\mathrm{km}$, and with a small best--worst spread, for example $0.0519$--$0.0548$ at $B{=}1\,\mathrm{km}$, indicating stable reconstruction quality.
Regarding the baseline predictors, Spectrum Surveying outperforms GP-DKL in the majority of cases, as its neural representation better captures complex non linear spatial dependencies. For example, under the proposed RL policy, Spectrum Surveying attains average RMSE of $0.1522/0.1218/0.0813$ for $B{=}1/2/4\,\mathrm{km}$, whereas GP-DKL remains substantially higher at $0.5187/0.4791/0.4217$. This gap reflects the difficulty of kernel based GP models in modeling urban complex radio map.

\begin{figure}[t]
    \centering
    \includegraphics[width=1\linewidth]{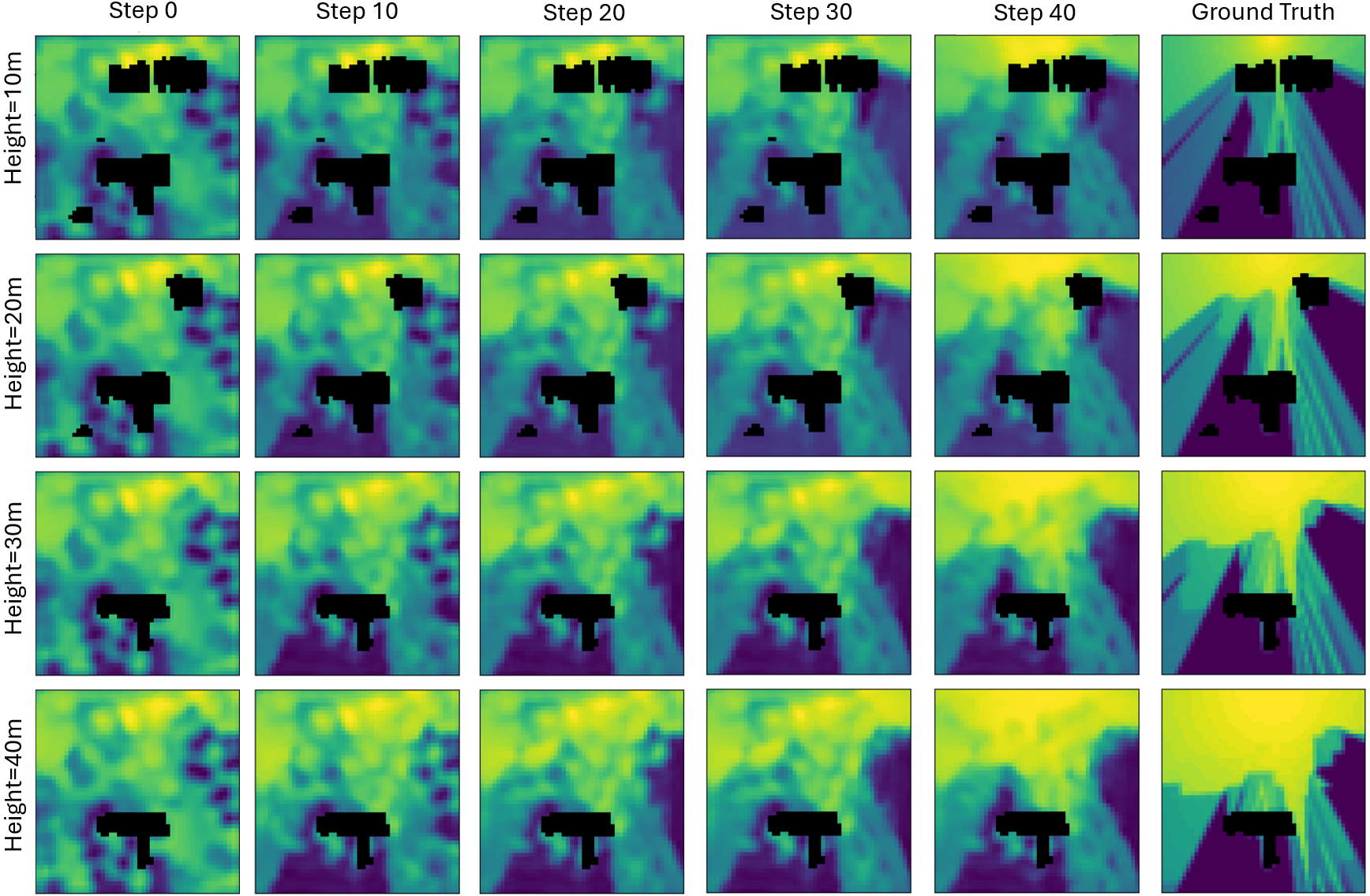}
    \caption{Reconstruction evolution over decision steps across altitudes under $B=4$, final RMSE 0.0363.}
    \label{fig:reconstruction_timeline}
\end{figure}

Beyond reconstruction back ends, Table~\ref{tab:baseline_matrix} also highlights the effect of exploration planning. The proposed RL policy consistently yields the lowest average and worst case RMSE compared with CMA-ES, MCTS, and random sampling for all three predictors, and the advantage is most pronounced under constrained budgets where measurement selection is critical. For instance, with the Bayesian U-Net at $B{=}1\,\mathrm{km}$, the proposed policy reduces average RMSE to $0.0536$ compared with $0.0694$ for CMA-ES, $0.0623$ for MCTS, and $0.0714$ for random sampling, while also achieving the lowest worst case RMSE. As the budget increases, all planners benefit from additional measurements; however, the proposed policy maintains the best accuracy and robustness, demonstrating higher sample efficiency. For qualitative insight, Fig.~\ref{fig:reconstruction_timeline} visualizes the reconstruction evolution produced by the proposed method at multiple decision steps across altitude layers. For clarity, only the proposed pipeline combining planner and reconstructor is plotted in this timeline; the relative advantages of different planners and reconstruction back ends are comprehensively quantified in Table~\ref{tab:baseline_matrix}.

\subsection{Runtime Evaluation}
The runtime of each module is evaluated with $H=W=50$, $Z=4$, $S=10$, and $m_k=20$. After 20 warm-up runs, the median latency over 100 repetitions is reported in Table~\ref{tab:runtime_breakdown}.

\begin{table}[!ht]
    \centering
    \caption{Module-Wise Runtime per Replanning Step.}
    \label{tab:runtime_breakdown}
    \resizebox{\columnwidth}{!}{%
    \begin{tabular}{lcc}
        \toprule
        Module & Complexity & Median latency \\
        \midrule
        Voxelization and map fusion & $\mathcal{O}(ZHW)$ & 0.010~ms \\
        PRM sampling and collision checking & $\mathcal{O}(m_k)$ & 0.849~ms \\
        Bayesian belief update ($S$ forward passes) & $\mathcal{O}(SZHW)$ & 28.54~ms \\
        CNN context encoding & $\mathcal{O}(ZHW)$ & 0.290~ms \\
        Transformer policy & $\mathcal{O}(m_k^2)$ & 0.566~ms \\
        \bottomrule
    \end{tabular}%
    }
\end{table}

The total median latency is approximately 30.26~ms per replanning step. Most of the runtime is associated with the $S$ stochastic forward passes used for the Bayesian belief update.

\subsection{Ablation Studies}

\subsubsection{Effect of the Observation-Mask Distribution}

\begin{figure}[!t]
    \centering
    \includegraphics[width=\columnwidth]{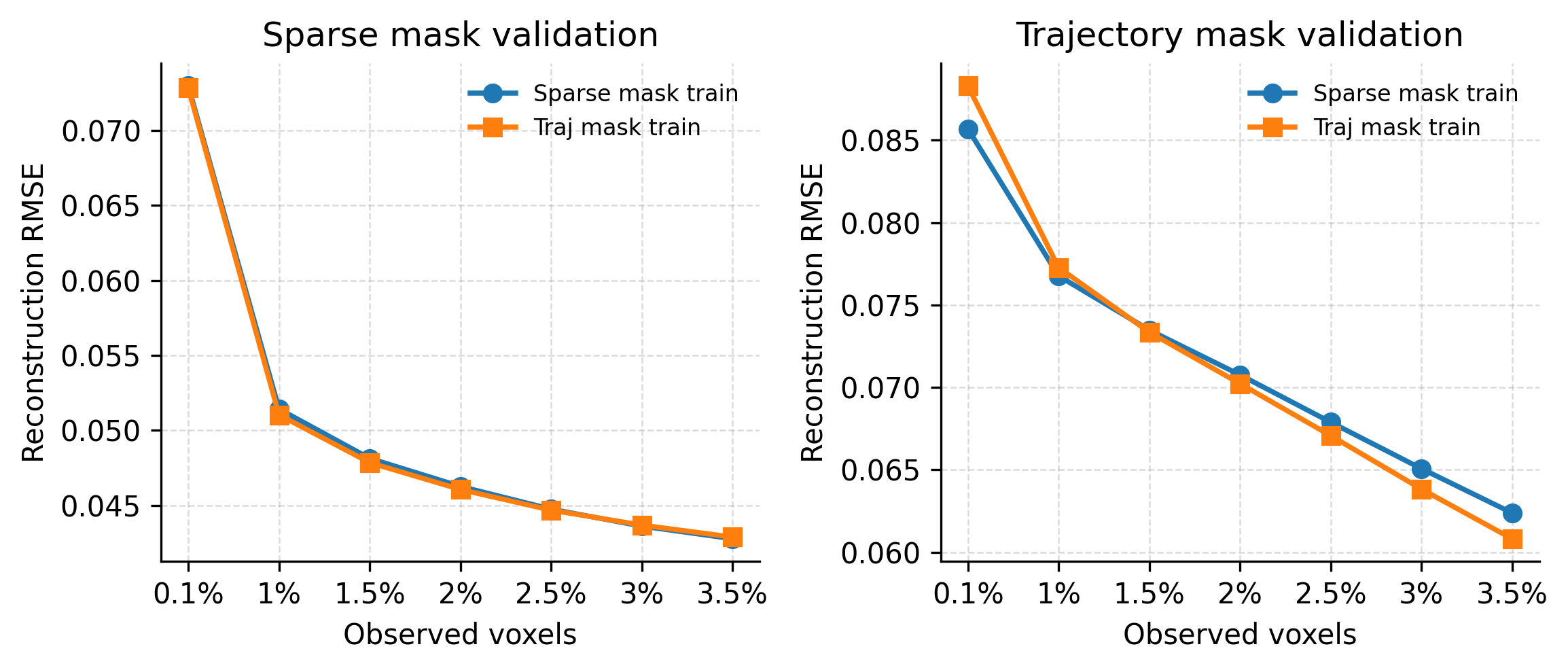}
    \caption{Stage-I RMSE for sparse-mask and trajectory-mask training on sparse-mask and trajectory-mask validation data.}
    \label{fig:two_mask_compare}
    \vspace{6pt}
    \centering
    \includegraphics[width=\columnwidth]{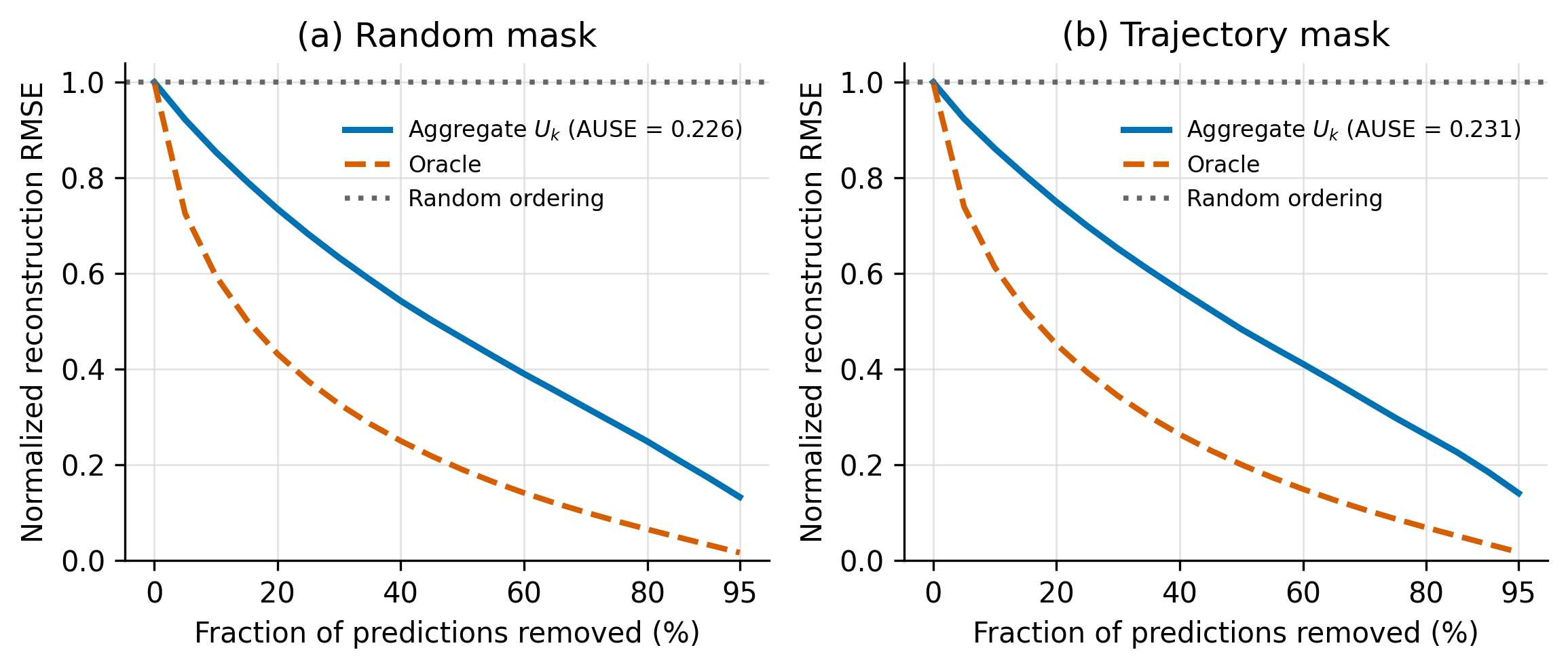}
    \caption{Uncertainty sparsification curves under random and trajectory observation masks, using the same frozen Stage-I model, test maps, and observation counts.}
    \label{fig:ause_mask}
    \vspace{8pt}
    \begin{minipage}{\columnwidth}
    \centering
    \captionof{table}{Effect of Stage-I Training Masks on Average End-to-End RMSE.}
    \label{tab:mask_end_to_end}
    \renewcommand{\arraystretch}{1.15}
    \resizebox{\columnwidth}{!}{%
    \begin{tabular}{lccc}
        \toprule
        Stage-I training mask & $B=1~\mathrm{km}$ & $B=2~\mathrm{km}$ & $B=4~\mathrm{km}$ \\
        \midrule
        Random sparse mask & 0.0536 & \textbf{0.0452} & 0.0355 \\
        Trajectory mask & \textbf{0.0528} & 0.0460 & \textbf{0.0349} \\
        \bottomrule
    \end{tabular}%
    }
    \end{minipage}
\end{figure}

To assess the mismatch between random training masks and trajectory-induced observations, we train two Stage-I models using random sparse masks and simulated UAV trajectory masks under otherwise identical settings. Both models are evaluated on both mask types across observed-voxel ratios. Fig.~\ref{fig:two_mask_compare} shows nearly identical RMSE on random-mask validation data and close performance on trajectory-mask validation data.

We further assess the error-ranking quality of $U_k$ using sparsification curves and the area under the sparsification error (AUSE)~\cite{Ilg2018Uncertainty}, where lower values indicate closer agreement with the oracle error ranking. Keeping the same frozen Stage-I model, test maps, and observation counts, we vary only the spatial arrangement of observations. As shown in Fig.~\ref{fig:ause_mask}, the AUSE increases only slightly from $0.226$ for random masks to $0.231$ for trajectory masks, suggesting that the error-ranking quality of $U_k$ is largely preserved under the tested change.

Finally, we train a separate Stage-II policy for each frozen Stage-I model using the same RL configuration. Table~\ref{tab:mask_end_to_end} shows that the average end-to-end RMSE differs by less than $1.8\%$, with neither training-mask distribution consistently superior.

\subsubsection{Effect of MC Dropout on Stage I}

\begin{figure}[t]
    \centering
    \includegraphics[width=\columnwidth]{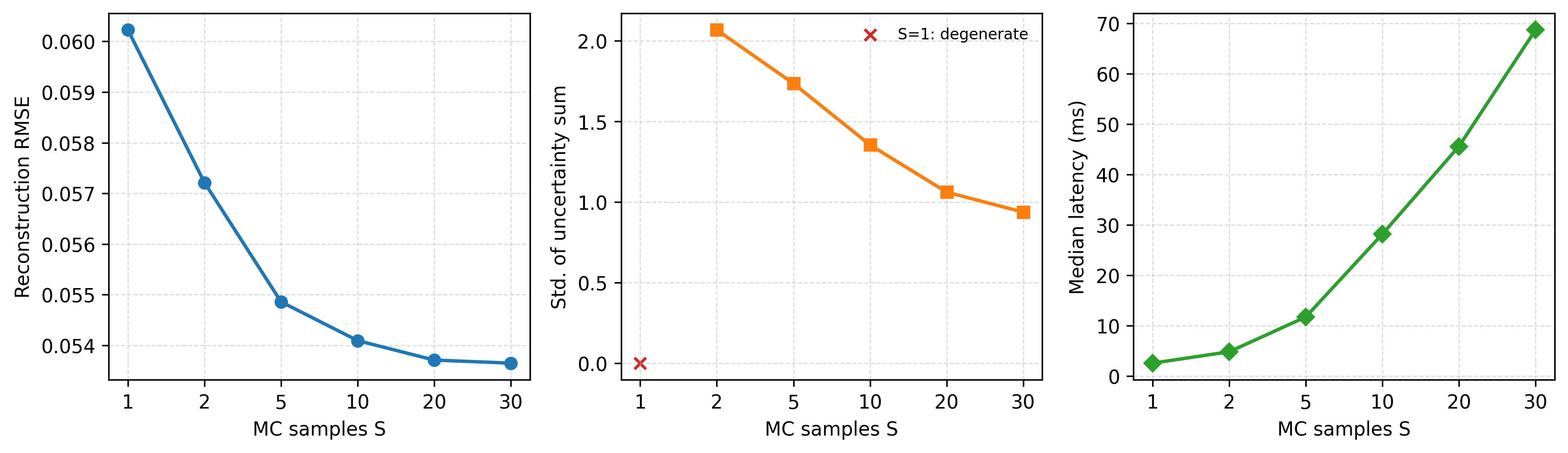}
    \caption{Effect of the number of MC Dropout samples on Stage-I reconstruction RMSE, aggregate-uncertainty stability, and median inference latency.}
    \label{fig:mc_sample}
\end{figure}

We evaluate the effect of the number of MC Dropout samples using $S\in\{1,2,5,10,20,30\}$ in terms of reconstruction RMSE, the standard deviation of the aggregate uncertainty, and inference latency. As shown in Fig.~\ref{fig:mc_sample}, increasing $S$ reduces the reconstruction error through ensemble averaging and stabilizes the uncertainty estimate. With $S=1$, the sample-based epistemic variance degenerates to zero. Beyond $S=10$, increasing $S$ to 20 and 30 reduces the RMSE by less than 1\% relative to $S=10$, while increasing the median inference latency by approximately 60\% and 145\%, respectively. The additional stochastic passes therefore incur a disproportionate computational cost for negligible accuracy gain. We use $S=10$ as the best accuracy--stability--latency trade-off.

\subsubsection{Effect of the Coefficient $\eta$ on Stage II}

The coefficient $\eta$ in Eq.~\eqref{eq:uncertainty_total} controls how strongly the aleatoric term contributes to the aggregate uncertainty $U_k$ relative to the epistemic term, and thus how the intrinsic reward trades off between noisy and unexplored regions. We retrain the policy with $\eta\in\{1,2,3,4,5\}$, keeping the frozen Stage-I predictor, $S=10$, $m_k=20$, and all PPO settings unchanged. Table~\ref{tab:ablation_eta} reports the resulting RMSE.

\begin{table}[t]
    \centering
    \caption{Effect of the Planning Coefficient $\eta$ on Stage II.}
    \label{tab:ablation_eta}
    \renewcommand{\arraystretch}{1.2}
    \resizebox{\columnwidth}{!}{%
    \begin{tabular}{lccc}
        \toprule
        Coefficient & $B=1$ & $B=2$ & $B=4$ \\
        \midrule
        $\eta=1$ (adopted) & \textbf{0.0536} & \textbf{0.0452} & \textbf{0.0355} \\
        $\eta=2$           & 0.0556 (+3.8\%) & 0.0453 (+0.1\%) & 0.0368 (+3.8\%) \\
        $\eta=3$           & 0.0559 (+4.3\%) & 0.0480 (+6.2\%) & 0.0395 (+11.2\%) \\
        $\eta=4$           & 0.0575 (+7.2\%) & 0.0471 (+4.2\%) & 0.0383 (+7.9\%) \\
        $\eta=5$           & 0.0571 (+6.5\%) & 0.0483 (+6.9\%) & 0.0390 (+9.9\%) \\
        \bottomrule
    \end{tabular}%
    }
\end{table}

Equal weighting achieves the lowest RMSE across all three budgets. Increasing $\eta$ overemphasizes largely irreducible observation noise, directing the agent toward noisy regions where additional measurements provide limited benefit and away from regions with high epistemic uncertainty. This degradation is most evident at $B=4$, where a larger budget compounds misdirected exploration.

\subsubsection{Effect of Policy Components on Stage II}

We retrain the agent with one policy component removed at a time, keeping all other settings and $m_k=20$ unchanged. Table~\ref{tab:ablation_policy} reports the resulting RMSE and the relative degradation.

\begin{table}[!t]
    \centering
    \caption{Ablation of Policy Components on Stage II.}
    \label{tab:ablation_policy}
    \renewcommand{\arraystretch}{1.2}
    \resizebox{\columnwidth}{!}{%
    \begin{tabular}{lccc}
        \toprule
        Variant & $B=1$ & $B=2$ & $B=4$ \\
        \midrule
        Full Transformer--PPO ($m_k=20$)  & \textbf{0.0536} & \textbf{0.0452} & \textbf{0.0355} \\
        w/o Fourier features              & 0.0542 (+1.1\%) & 0.0475 (+5.0\%) & 0.0361 (+1.7\%) \\
        w/o terminal uncertainty penalty  & 0.0546 (+1.9\%) & 0.0458 (+1.3\%) & 0.0357 (+0.7\%) \\
        Aleatoric uncertainty only        & 0.0545 (+1.7\%) & 0.0466 (+3.0\%) & 0.0360 (+1.3\%) \\
        Epistemic uncertainty only        & 0.0572 (+6.7\%) & 0.0509 (+12.7\%) & 0.0407 (+14.6\%) \\
        \bottomrule
    \end{tabular}%
    }
\end{table}

\begin{table}[!t]
    \centering
    \caption{Effect of the Candidate-Set Size $m_k$ on Stage II.}
    \label{tab:ablation_mk}
    \renewcommand{\arraystretch}{1.2}
    \resizebox{\columnwidth}{!}{%
    \begin{tabular}{lccc}
        \toprule
        Candidate number & $B=1$ & $B=2$ & $B=4$ \\
        \midrule
        $m_k=10$          & 0.0572 (+6.8\%) & 0.0462 (+2.3\%) & 0.0361 (+1.7\%) \\
        $m_k=20$ (adopted) & 0.0536 & 0.0452 & 0.0355 \\
        $m_k=30$          & 0.0525 ($-2.1\%$) & 0.0442 ($-2.3\%$) & 0.0345 ($-2.7\%$) \\
        $m_k=40$          & 0.0523 ($-2.5\%$) & 0.0439 ($-2.8\%$) & 0.0344 ($-3.0\%$) \\
        \bottomrule
    \end{tabular}%
    }
\end{table}

Removing the Fourier mapping costs $1.1\%$--$5.0\%$, indicating that the multi-frequency lifting of the waypoint coordinates helps the policy discriminate spatially adjacent candidates. Discarding the terminal penalty $-\beta\Sigma_k$ costs $0.7\%$--$1.9\%$, as the agent then optimizes only per-step uncertainty reduction and becomes myopic near budget exhaustion. The uncertainty source matters most: the aleatoric-only variant loses $1.3\%$--$3.0\%$ because the intrinsic noise is largely irreducible and thus a weak exploration signal, while the epistemic-only variant is the worst, degrading by $6.7\%/12.7\%/14.6\%$ for $B=1/2/4$. The gap widens with the budget since model uncertainty alone drives the UAV toward unvisited but not necessarily informative regions. The aggregate uncertainty combining both terms therefore yields the best accuracy.

\subsubsection{Effect of the Candidate-Set Size on Stage II}

The cardinality $m_k$ controls the resolution of the discretized action space. We retrain the policy with $m_k\in\{10,20,30,40\}$ under otherwise identical settings, as summarized in Table~\ref{tab:ablation_mk}.

\begin{table}[!t]
    \centering
    \caption{Configuration of the Real-World Measurement Campaign.}
    \label{tab:field_config}
    \renewcommand{\arraystretch}{1.2}
    \begin{tabular}{lc}
        \toprule
        {Parameter}                & {Value}                          \\
        \midrule
        \multicolumn{2}{c}{\textbf{Measurement Platform}}             \\
        \midrule
        UAV Platform               & DJI Matrice 350 RTK              \\
        Spectrum Scanner           & R\&S TSMA6                       \\
        Flight Altitude            & 40~m AGL                         \\
        Coverage Footprint         & $300 \times 200~\mathrm{m}^2$    \\
        \midrule
        \multicolumn{2}{c}{\textbf{Radio Configuration}}              \\
        \midrule
        NR-ARFCN                   & 504990                           \\
        Center Frequency           & 2524.95~MHz (n41)                \\
        SSB\_RP Range      & $-120.0$ to $-49.9$~dBm          \\
        \midrule
        \multicolumn{2}{c}{\textbf{Collected Dataset}}                \\
        \midrule
        Valid Measurement Records  & 742988                        \\
        Detected Cells             & 45 PCIs                          \\
        Beam Indices               & 0--7                             \\
        Distinct (PCI, Beam) Maps  & 180                              \\
        \midrule
        \multicolumn{2}{c}{\textbf{Evaluated Map}}      \\
        \midrule
        Evaluated Cell / Beam      & PCI 657 / Beam 6                 \\
        Records / Positions        & 12004 / 2390               \\
        SSB\_RP Range              & $-106.9$ to $-73.4$~dBm          \\
        Evaluation Grid            & $50 \times 50$                    \\
        Route Length               & 385~m                            \\
        \bottomrule
    \end{tabular}
\end{table}

With $m_k=10$, the coarse action space increases the RMSE by up to $6.8\%$ because the roadmap may miss informative waypoints. Increasing $m_k$ from $20$ to $30$ or $40$ reduces the RMSE by only $2.1\%$--$3.0\%$, with only marginal further reductions from $30$ to $40$. Meanwhile, PRM sampling and Transformer self-attention scale as $\mathcal{O}(m_k)$ and $\mathcal{O}(m_k^2)$, respectively. We therefore adopt $m_k=20$ as a balance between accuracy and complexity.

\subsection{Small-Scale Real-World Field Test}

\begin{figure}[!t]
    \centering
    \includegraphics[width=\columnwidth]{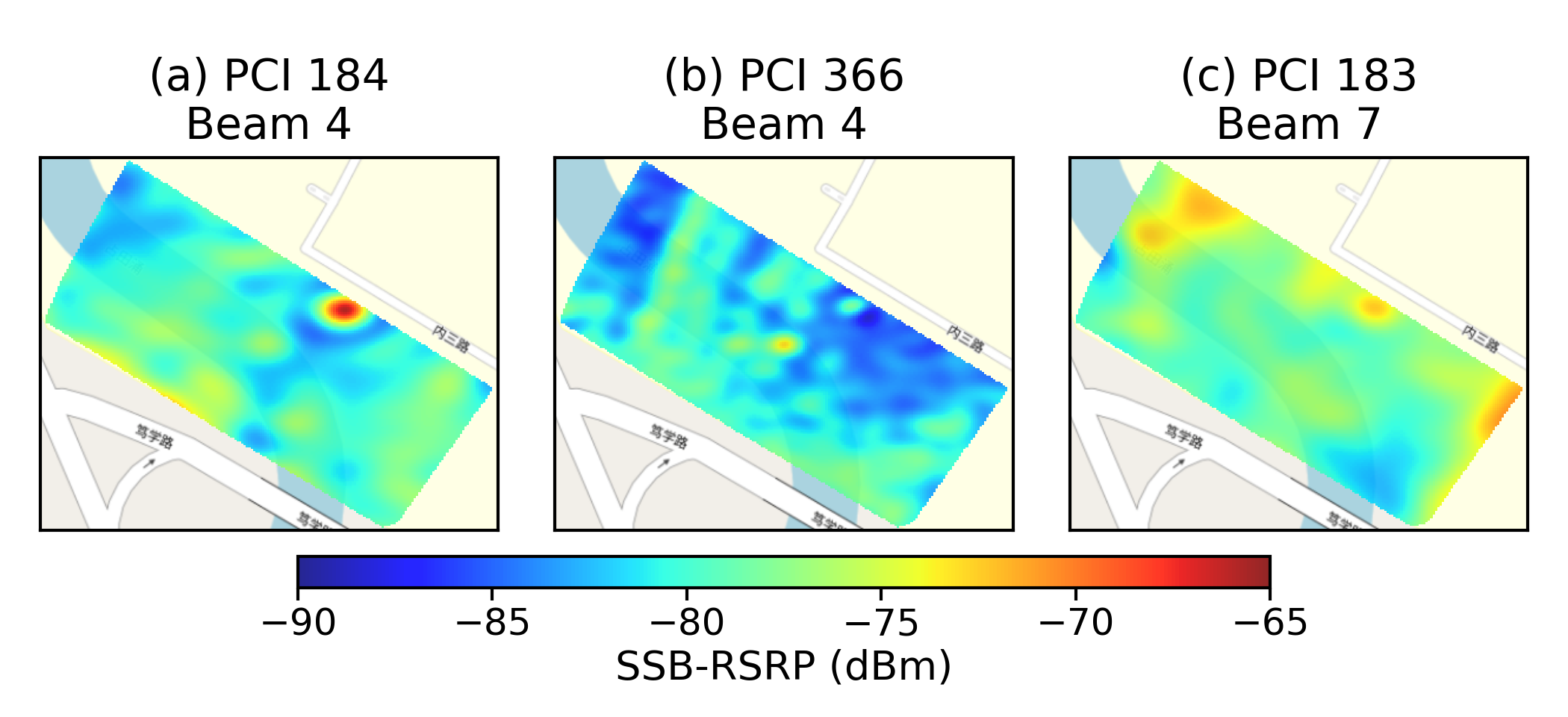}
    \caption{Radio map examples from the real-world dataset.}
    \label{fig:field_dataset_pcis}
\par\vspace{8pt}

    \centering
    \begin{subfigure}[t]{0.3\linewidth}
        \centering
        \includegraphics[width=\linewidth]{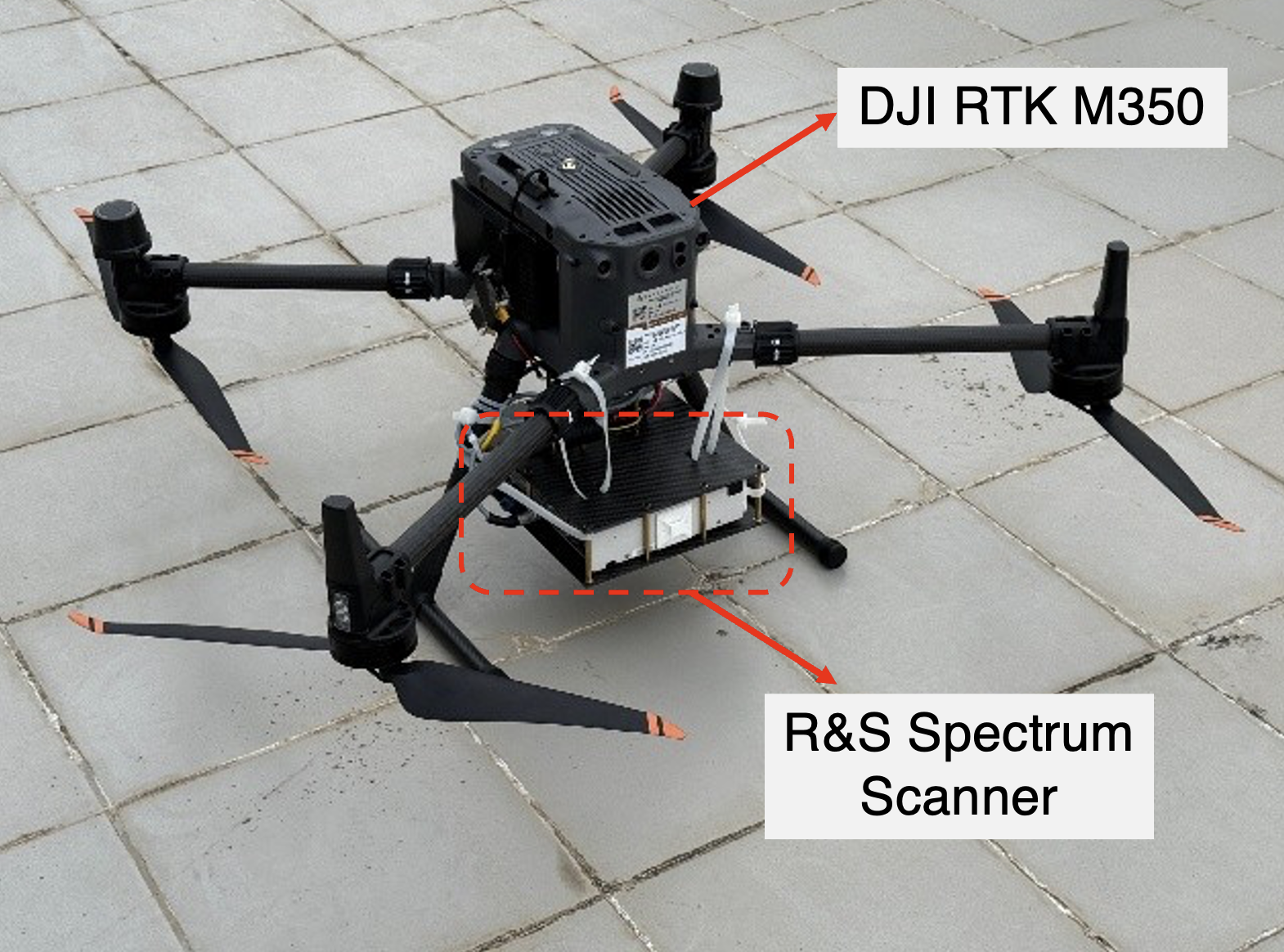}
        \caption{Measurement tool.}
        \label{fig:field_uav_platform}
    \end{subfigure}%
    \hspace{0.06\linewidth}%
    \begin{subfigure}[t]{0.3\linewidth}
        \centering
        \includegraphics[width=\linewidth]{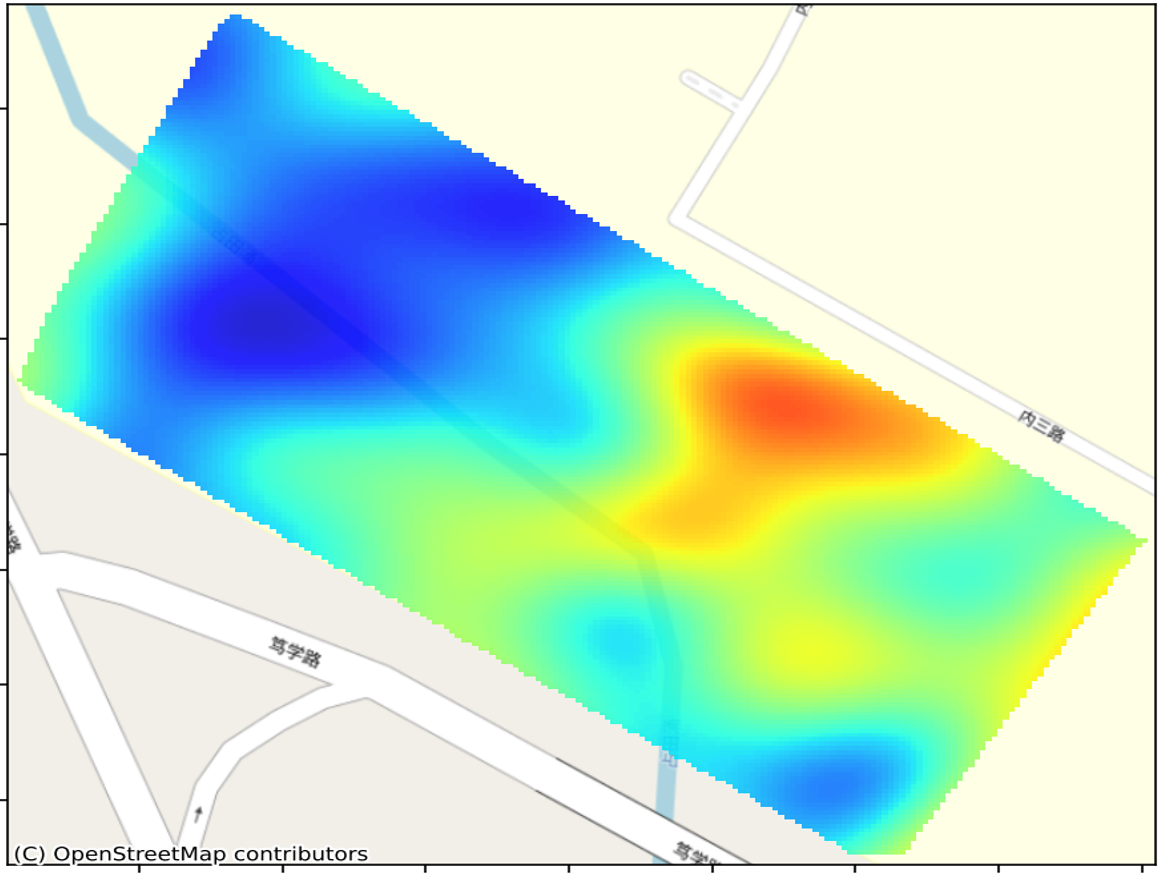}
        \caption{Real aerial radio map.}
        \label{fig:field_real_gt}
    \end{subfigure}
    
    \begin{subfigure}[t]{0.3\linewidth}
        \centering
        \includegraphics[width=\linewidth]{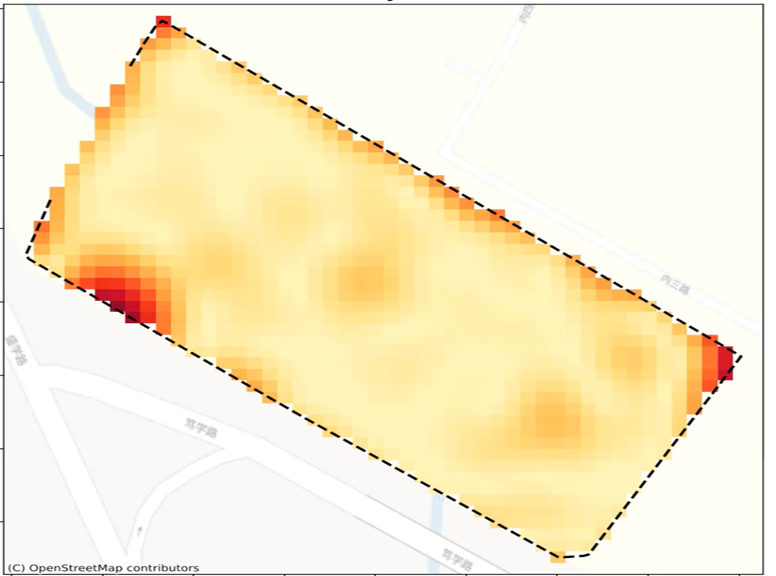}
        \caption{Uncertainty map.}
        \label{fig:field_vis_unc_real}
    \end{subfigure}%
    \hspace{0.04\linewidth}%
    \begin{subfigure}[t]{0.32\linewidth}
        \includegraphics[width=\linewidth]{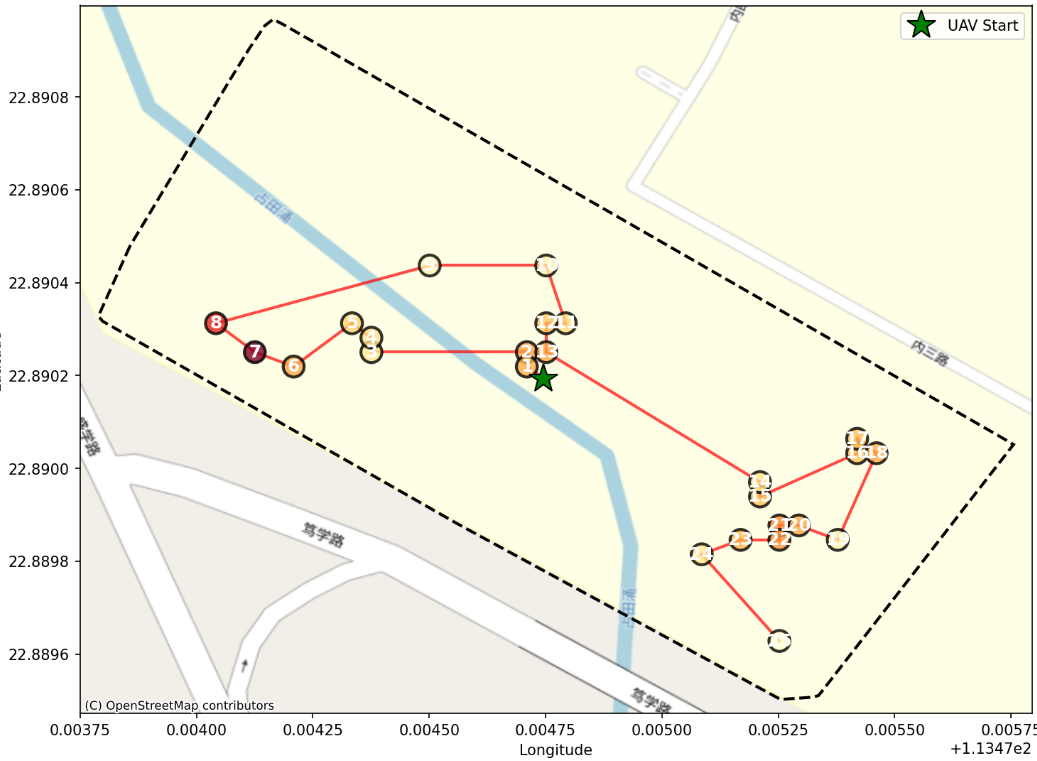}
        \caption{Trajectory visualization.}
        \label{fig:field_uav_traj}
    \end{subfigure}
    \vspace{-0.1cm}
    \caption{Offline real-world field test configuration in HKUST-GZ.}
    \label{fig:field_real_uav}
\end{figure}

The measurement campaign was conducted over the HKUST(GZ) campus, within a 300m$\times$200m operating area, as illustrated in Fig.~\ref{fig:field_real_uav}. A DJI Matrice 350 RTK carrying an R\&S TSMA6 spectrum scanner flew at a constant altitude of $40~\mathrm{m}$ AGL for approximately 52 minutes, logging SSB-based 5G NR measurements at NR-ARFCN 504990 ($2524.95~\mathrm{MHz}$, band n41). Since the scanner decodes every detectable SSB in the band, one flight yields a separate radio map per (PCI, beam index) pair: the campaign produced 742988 valid records covering 45 physical cells and 8 SSB beams. Table~\ref{tab:field_config} summarizes the configuration, while Fig.~\ref{fig:field_dataset_pcis} shows radio map examples for three densely sampled PCI--beam pairs from this dataset.

The Stage-I reconstructor is first pre-trained on the Sionna ray tracing dataset and then fine-tuned on the remaining maps collected during the campaign. To prevent data leakage, the evaluated map (PCI 657, beam 6) is held out throughout fine-tuning. Reconstruction error is computed over the 1473 cells within the convex hull of the flight footprint on a $50\times50$ grid. Given the initial measured samples, the fine-tuned Bayesian U-Net produces a provisional reconstruction and an uncertainty map. Based on this uncertainty map, several high-uncertainty regions are selected to form a single refinement trajectory. The trajectory is executed without intermediate replanning or waypoint re-selection. The flight lasts approximately $10$ min and covers a total distance of $385~\mathrm{m}$.

This field test examines whether the Stage-I uncertainty estimates remain useful for guiding sampling with real RSRP measurements. Under the same $385~\mathrm{m}$ travel budget, the uncertainty-guided trajectory reduces the reconstruction RMSE from $0.175$, achieved by the predefined Lawnmower sweep, to $0.082$. This result serves only as a sanity check: additional measurements in regions identified as uncertain improve reconstruction, suggesting that the learned uncertainty can support offline trajectory refinement. Because measurements are collected at a single altitude, the experiment evaluates the Stage-I estimator only on a 2D slice; end-to-end real-world validation of the complete 3D-URAM framework remains future work.

\section{Conclusions}
This paper presented the first solution for actively unveiling 3D air-ground radio maps in unknown environment. The proposed 3D-URAM framework began by training a Bayesian U-Net via dual masked inpainting, where sparse measurements and geometric priors were sequentially mapped to a continuous signal field with spatially resolved predictive uncertainty. This belief state then guided a Transformer-based policy on a dynamic probabilistic roadmap to maximize long horizon information gain under strict travel budgets. The approach substantially outperformed state of the art baselines in both reconstruction accuracy and sample efficiency. Furthermore, a supplementary field measurement case study suggested that the learned uncertainty also supported offline batch trajectory refinement on real-world measurements. Future work includes broader field validation and incorporation of cross frequency prediction.

\ifCLASSOPTIONcaptionsoff
    \newpage
\fi

\bibliographystyle{IEEEtran}
\bibliography{references}

\end{document}